\documentclass[onefignum,onetabnum]{siamonline220329}

\newcommand{\jfpackage}[1] {\usepackage{./#1}}

\newcommand{\indef}[1] {\input{./#1}}
\jfpackage{jf-all}

\definecolor{ForestGreen}{RGB}{34,139,34}

\usepackage{upgreek} 
\usepackage{hyperref}

\usepackage{cite}
\usepackage{booktabs}
\usepackage{enumerate}

\usepackage{array} 
\usepackage{xcolor} 
\usepackage{colortbl} 

\usepackage{algorithm} 
\usepackage{algpseudocode}

\newcommand{\codo}[1] {}
\long\def\red#1{\bgroup\color{red}#1\egroup}

\usepackage{graphicx} 
\usepackage{./support-caption}
\usepackage{subcaption}

\newcommand{\wiki}[1]	{\href{#1}{[wiki]}}

\newcommand{\tref}[1]	{Theorem~\ref{#1}}

\newcommand{\tofref}[1]	{\todoa{Fig. (XXX)}}

\newcommand{\todof}[1]	{\todoa{Fig. (XXX)}}

\renewcommand{\defequ}	{\xmath{\triangleq}}

\newcommand{\arrow}	{\inmath{\rightarrow}}

\newcommand{\vsig}	{\blmath{\sigma}}

\newcommand{\vs}	{\jfunop{\bm{s}}}

\newcommand{\w}	{\blmath{w}}
\newcommand{\vone}	{\blmath{1}}

\newcommand{\bDel}	{\blmath{\Delta}}

\newcommand{\subrm}[1]	{_{\mathrm{#1}}}

\renewcommand{\reals}	{\jfunl{\mathbb{R}}}
\newcommand{\complex}	{\jfunl{\mathbb{C}}}

\newcommand{\ints}	{\jfunl{\mathbb{Z}}}

\newcommand{\A}	{\blmath{A}}
\newcommand{\B}	{\blmath{B}}

\newcommand{\bE}	{\blmath{E}}

\newcommand{\G}	{\blmath{G}}
\renewcommand{\H}	{\blmath{H}}
\newcommand{\I}	{\blmath{I}}

\renewcommand{\P}	{\blmath{P}}

\newcommand{\bS}	{\blmath{S}}
\newcommand{\T}	{\blmath{T}}
\newcommand{\U}	{\blmath{U}}
\newcommand{\V}	{\blmath{V}}
\newcommand{\W}	{\blmath{W}}

\newcommand{\X}	{\blmath{X}}
\newcommand{\Xh}	{\X^`}
\newcommand{\Y}	{\blmath{Y}}

\newcommand{\norminf}[1]	{\xmath{\norm{#1}_{\infty}}}

\newcommand{\mnorm}[1]	{\norm{#1}}

\newcommand{\mnorms}[1]	{\| #1 \|} 

\newcommand{\mnormfrob}[1]	{\xmath{\mnorm{#1}\subrm{F}}}
\newcommand{\mnormsfrob}[1]	{\xmath{\mnorms{#1}\subrm{F}}}

\newcommand{\reg}	{\jfunl{\upbeta}}

\newcommand{\jcal}[1]	{\jfunl{\mathcal{#1}}}

\newcommand{\cA}	{\jcal{A}}

\newcommand{\vp}	{\blmath{p}}

\newcommand{\vu}	{\blmath{u}}
\newcommand{\vv}	{\blmath{v}}
\newcommand{\y}	{\blmath{y}}

\newcommand{\xsym}	{x}

\newcommand{\x}	{\blmath{\xsym}}

\newcommand{\nLsym}	{\ooalign{$\mathsf{L}$\cr\kern+0.0em\textsf{-}}}

\newcommand{\potsym}	{\inmath{\psi}}
\newcommand{\pot}	{\jfunop{\potsym}}

\newcommand{\dpot}	{\jfunop{\potsym}^.}

\newcommand{\wpot}	{\jfunop{\omega}_{\potsym}}

\newcommand{\kostsym}	{\inmath{\Psi}}
\newcommand{\kost}	{\jfunop{\kostsym}}

\renewcommand{\L} {\blmath{L}}

\newcommand{\MbyN} {\xmath{M \times N}}

\newcommand{\CMN} {\xmath{\complex^{\MbyN}}}

\newcommand{\CMM} {\xmath{\complex^{M\times M}}}
\newcommand{\CSN} {\xmath{\complex^{S\times N}}}

\newcommand{\CPN} {\xmath{\mathbb{C}^{P\times N}}}

\newcommand{\RMN} {\xmath{\reals^{\MbyN}}}

\renewcommand{\S}{\bS} 

\newcommand{\fpot} {\xmath{f_{\pot}}}
\newcommand{\fpotc} {\xmath{f_{\pot, \bm w}}}
\newcommand{\fpotw} {\xmath{f_{\pot, \bm w}}}
\newcommand{\rpot} {\xmath{R_{\pot}}}
\newcommand{\rpotk} {\xmath{R_{\pot, K}}}
\newcommand{\rpotc} {\xmath{R_{\pot, \bm w}}}
\newcommand{\rpotw} {\xmath{R_{\pot, \bm w}}}
\newcommand{\rpotwk} {\xmath{R_{\pot, \bm w_K}}}
\newcommand{\hpot} {\xmath{h_{\pot}}}
\newcommand{\hpotw} {\xmath{h_{\pot, \w}}}

\newcommand{\qpotw} {\xmath{Q_{\bm w}}}
\newcommand{\gpotw} {\xmath{G_{\pot, \bm w}}}
\newcommand{\gpot} {G_{\pot}}

\newcommand{\sumkm}[1][k]	{\sum_{#1=1}^M}

\providecommand{\rinprod}[2]	{\xmath{\mathop{\real{\langle #1,\, #2 \rangle}}\nolimits}}
\providecommand{\rInprod}[2]	{\xmath{\real{\left\langle #1,\ #2 \right\rangle}}}

\newcommand{\potdot} {\xmath{\pot\!.}}
\newcommand{\dpotdot}[1] {\xmath{\dpot\!.(#1)}}
\newcommand{\wpotdot}[1] {\xmath{\wpot\!.(#1)}}

\newcommand{\gpotL} {\xmath{\gpot^{\mathrm{L}}}}
\newcommand{\gpotR} {\xmath{\gpot^{\mathrm{W}}}}
\newcommand{\alfs} {\xmath{\bar{\alf}}}

\newcommand{\gpotwL} {\xmath{\gpotw^{\mathrm{L}}}}
\newcommand{\gpotwR} {\xmath{\gpotw^{\mathrm{W}}}}
\newcommand{\gpotwkL} {\xmath{G_{\pot, \w_K}^{\mathrm{L}}}}
\newcommand{\gpotwkR} {\xmath{G_{\pot, \w_K}^{\mathrm{W}}}}

\newcommand{\Pkk}{\xmath{\P_{k_1,k_2}}}

\newsiamthm{remark}{Remark}

\title{
Smooth optimization 
using global and local low-rank regularizers  
}

\date{\today}

\author{Rodrigo~A.~Lobos%
\footnotemark[2]
\thanks{Department of Biomedical Engineering, University of Michigan, Ann Arbor, MI, 48109
(\email{rlobos@umich.edu})}
\and Javier Salazar Cavazos%
\thanks{Department of Electrical Engineering and Computer Science,
University of Michigan, Ann Arbor, MI, 48109
(\email{javiersc@umich.edu}, \email{rajnrao@umich.edu}, \email{fessler@umich.edu})
\funding{
Research supported in part by
NIH Grants
R01 EB023618,
U01 EB026977,
R01 EB035618, 
R21 EB034344, 
and
R21 AG061839,
NSF grant 2331590.
}
}
\and Raj Rao Nadakuditi\footnotemark[2] 
\and Jeffrey~A.~Fessler\footnotemark[2]
}

\begin{document}

\maketitle

\uncomment
{
\begin{abstract}%

Many inverse problems and signal processing problems
involve low-rank regularizers based on the nuclear norm.
Commonly, proximal gradient methods (PGM) are adopted
to solve this type of non-smooth problems
as they can offer fast and guaranteed convergence.
However, PGM methods cannot be simply applied in settings
where low-rank models are imposed locally on overlapping patches;
therefore, heuristic approaches have been proposed that
lack convergence guarantees.
In this work we propose to replace the nuclear norm with a smooth approximation
in which a Huber-type function is applied to each singular value.
By providing a theoretical framework based on singular value function theory,
we show that important properties can be established for the proposed regularizer,
such as:
convexity, differentiability, and Lipschitz continuity of the gradient.
Moreover, we provide a closed-form expression
for the regularizer gradient,
enabling the use of standard iterative gradient-based optimization algorithms
(e.g., nonlinear conjugate gradient)
that can easily address the case of overlapping patches
and have well-known convergence guarantees.
In addition, we provide a novel step-size selection strategy
based on a quadratic majorizer
of the line-search function
that leverages the Huber characteristics of the proposed regularizer.
Finally, we assess the proposed optimization framework
by providing empirical results in dynamic magnetic resonance imaging (MRI) reconstruction
in the context of local low-rank models with overlapping patches.
\end{abstract}
}

\section{Introduction}
Some inverse problems
and many signal processing problems
involving low-rank models
are formulated
as optimization problems
using global low-rank regularizers
as follows%
\footnote{Throughout the paper we use capital and lower case bold letters
to denote matrices and vectors, respectively.
The entry in the $k$th row and $l$th column of a matrix \X
is denoted either as $X_{kl}$ or $[\X]_{kl}$. 
Analogously, the $k$th entry of a vector \x
is denoted either as $x_k$ or $[\x]_k$.
Besides the nuclear norm,
in later derivations we also use the Frobenius norm for matrices,
denoted by $\mnormfrob{\cdot}$.
For vectors, we denote by
$\norm{\cdot}$ and $\norminf{\cdot}$ the
$\ell_2$-norm and the $\ell_{\infty}$-norm, 
respectively.}: 
\be
\Xh \in \argmin{\X \in \CMN}
f(\X) + \reg \mnorm{\X}_*
,\ee{e,kost}%
where $f$ typically corresponds to a function that measures data consistency; 
$\mnorm{\cdot}_*$ denotes the nuclear norm;
and \reg is a regularization parameter.
Example applications include
robust PCA
\cite{candes:11:rpc},
matrix completion
\cite{davenport:16:aoo,chi:18:lrm},
parallel MRI
\cite{
jin:16:agf, 
zhang:20:irw,
zhang:22:amr 
},
low-rank + sparse models
for dynamic MRI
\cite{otazo:15:lrp},
among many others
\cite{
udell:16:glr,
chen:18:hsi,
wen:18:aso,
chi:19:nom,
jacob:20:slr,
haldar:20:lpi,
zha:23:lns
}.

In the usual case where $f$ has a Lipschitz continuous gradient,
these optimization problems
are solved easily
using proximal gradient methods
like the
proximal optimized gradient method (POGM)
\cite{taylor:17:ewc-composite,kim:18:aro}, as the nuclear norm is non-smooth
but ``prox friendly.''
Empirical results in dynamic MRI
show that POGM converges faster
than FISTA for such cost functions
\cite{lin:19:edp}.
Nevertheless,
these first-order (accelerated) proximal gradient methods
have worst-case $O(1/k^2)$ convergence rates
that can be slow
compared to optimization methods
for smooth problems \cite{riahi:22:otc,das2024nonlinear}.
Augmented Lagrangian and ADMM algorithms
are another option,
but these require additional tuning parameters
and do not have faster convergence rates.
Additional limitations for proximal gradient methods
are encountered when using local low-rank models,
where low-rank characteristics are promoted locally on overlapping patches constructed from \X.
Applications include:
matrix approximation
\cite{lee:13:llr},
MRI reconstruction \cite{trzasko2011calibrationless, saucedo2017improved},
dynamic MRI
\cite{
trzasko:11:lvg,
ke:21:llr,
cruz2023low
},
multi-contrast and quantitative MRI
\cite{
yaman:17:llr, 
tamir:17:tss 
},
MR denoising
\cite{lu:22:ann},
fMRI denoising
\cite{
vizioli:21:ltt, 
comby:23:dof, 
meyer::ect 
},
fMRI dynamic image reconstruction
\cite{guo:20:hro},
and MR motion correction
\cite{chen:23:isl}. 
As an example,
in dynamic MRI reconstruction the optimization problem has the form%
\footnote{Here we briefly describe the dynamic MRI reconstruction optimization problem
using local low-rank regularizers.
Section \ref{sec,mri} provides more details.
}:
\be
\Xh \in \argmin{\X \in \CMN}
f(\X) + \reg \sum_{\vs \in \Lambda}\sum_{\vp \in \Gamma}\mnorm{\jcal{P}_{\vp}(\jcal{S}_{\vs}(\X))}_*
,\ee{e,kost_local}
where $\Lambda, \Gamma \subseteq \ints^d$ are sets of shift and location indexes, respectively,
and $d$ corresponds to the image voxel dimensions%
\footnote{For simplicity we consider $d=2$ throughout the paper (\ie, 2D images);
however, our results can be easily extended to higher dimensions.};
$\jcal{S}_{\vs}:\CMN\arrow \CMN$ is a linear operator
that circularly shifts the images composing the columns of \X
according to the shift index \vs;
and $\jcal{P}_{\vp}:\CMN\arrow \CPN$ is a linear operator
that constructs a matrix from a patch extracted from \X
whose entries depend on the location index $\vp$.
In many applications these patches are considered overlapping (\ie, share entries of \X),
which makes standard proximal algorithms unsuitable for solving \eref{e,kost_local},
as there is no known simple proximal mapping for the regularizer in this case.
This challenge has led to various heuristics
like ``cycle spinning''
\cite{trzasko:17:spd}, 
proximal averaging
\cite{bauschke:08:tpa},
and other approaches
that do not have any convergence guarantees
\cite{zhang:15:apm}.
Replacing the nuclear norm with a \emph{smooth approximation}
could facilitate the use
of 
standard iterative gradient-based optimization algorithms
like nonlinear conjugate gradient (NCG),
or perhaps even
second-order optimization methods
like quasi-Newton methods.
These algorithms might lead to faster convergence while providing convergence guarantees.
Moreover, they could be applied directly to the local low-rank formulation
even in the case of overlapping patches. 

The nuclear norm in
\eref{e,kost}
is a convex relaxation
of a regularizer
based on the rank of \X,
so it is already a kind of approximation
to the ``ideal'' low-rank model.
Thus, it seems quite reasonable
to entertain further approximations.
This work investigates the use of smooth regularizers
instead of non-smooth alternatives such as the nuclear norm.
Specifically, we study regularizers of the form:
\be
\rpot(\X) \defequ \sum_{k=1}^r \pot(\sig_k(\X))
,\ee{e,reg_smooth}
where $r=\min(M,N)$; $\sig_k(\X)$ denotes the $k$th singular value of $\X$
and \pot is a smooth function subject to mild regularity conditions
that can be either convex or non-convex.
We study the case where \pot satisfies the so-called Huber conditions \cite[p. 184]{huber:81},
which is why we call \rpot a \textit{Huber-based low-rank regularizer}.
Then, we propose to impose global low-rank models by solving: 
\be
\Xh \in \argmin{\X \in \CMN}
\kost_{\text{Global}}(\X)
,\quad
\kost_{\text{Global}}(\X) =
f(\X) + \reg \rpot(\X)
,\ee{e,kost_global_h}
and to impose local low-rank models by solving:
\be
\Xh \in \argmin{\X \in \CMN}
\kost_{\text{Local}}(\X)
,\quad
\kost_{\text{Local}}(\X) =
f(\X) + \reg \sum_{\vs \in \Lambda}\sum_{\vp \in \Gamma}\rpot(\jcal{P}_{\vp}(\jcal{S}_{\vs}(\X)))
.\ee{e,kost_local_h}
The local case with overlapping patches
is our primary motivation,
since that is the case where the regularizer
is not ``prox friendly.''

Similar low-rank regularizers to \eref{e,reg_smooth}
have been proposed in the global  and local  cases
in \cite{lu2014generalized} and \cite{liu2020multiplicative}, respectively.
The resulting optimization problem 
is solved by approximating the regularizer 
using a first-order expansion
that produces a non-convex and non-smooth cost function. 
This setting allows one to establish algorithms with convergence guarantees
as shown in \cite{lu2014generalized} and \cite{liu2020multiplicative}.
However, in the local low-rank case,
the algorithm proposed in \cite{liu2020multiplicative} 
introduces tuning parameters that affect convergence speed,
and the total number of these parameters grows with the number of patches. 
In contexts like dynamic MRI reconstruction
the cost function \eref{e,kost_local_h} might include hundreds of patches, 
hindering the applicability of algorithms like the one in \cite{liu2020multiplicative}. 
Non-convex Schatten-$p$ quasi-norms
with $0 < p < 1$
have also been investigated
as beneficial alternatives to the nuclear norm,
\eg,
\cite{ongie:17:afa}. 
Other non-convex functions of the singular values
have also been investigated,
\eg,
\cite{oh2016partial,cavazos2025alpcah,haldar:14:lrm, chen2013reduced}.

The regularizer proposed in \eref{e,reg_smooth}
corresponds to a unitarily invariant function \cite{lewis1995convex},
and therefore it only depends on the singular values.
Functions with these characteristics are known as singular value functions
\cite{lewis2005nonsmooth,lewis2005nonsmoothII},
and their properties
(\eg, convexity, differentiability, Lipschitz continuity of the gradient)
have been extensively investigated over the last two decades
\cite{lewis1995convex, lewis1996convex, lewis2001twice, lewis2002quadratic, qi2003semismoothness,
lewis2005nonsmooth,lewis2005nonsmoothII,sendov2006generalized,sendov2007higher}.
However, to the best of our knowledge,
the application of these properties has not been explored
in inverse problems using low-rank models.
In this work we show that relevant properties for the optimization of \rpot
are directly related to the analogous properties of \pot.
Using these derivations,
we show that it is possible to use standard gradient-based iterative optimization algorithms
(\eg, NCG, quasi-Newton methods)
to solve \eref{e,kost_global_h} and \eref{e,kost_local_h},
that offer convergence guarantees 
without needing to introduce tuning parameters.

Iterative gradient-based optimization algorithms usually require two main computations:
the cost function gradient,
and a step size in each iteration that guarantees a decrease of the cost function.
We show that the gradient of the cost function can be calculated in closed form
by relying on singular value function theory to calculate $\nabla\rpot(\X)$.
Then we provide a step-size selection strategy
that exploits the Huber/Lipschitz characteristics of \pot.
Specifically, 
we propose a majorize-minimize (MM) dynamic step-size selection strategy
to monotonically decrease the line search function at each iteration,
where a quadratic majorizer is constructed 
by leveraging on the Huber conditions satisfied by \pot.
This construction is based on deriving novel majorizers for the proposed regularizer \rpot,
which is one of our main theoretical contributions in this work.

In addition, we generalize our theoretical framework
by considering a weighted version of the proposed regularizers,
that we call \textit{weighted Huber-based low-rank regularizers}.
Specifically, we study regularizers of the form:
\be
\rpotc(\X) \defequ \sum_{k=1}^r w_k\pot(\sig_k(\X)),
\ee{e,reg_weight}
where $\w =[w_1, \ldots , w_r]^T$ corresponds to a nonzero vector of nonnegative weights,
and $(\cdot)^T$ denotes the transpose operation.
To properly impose low-rank models,
\rpotw should penalize more small singular values than large singular values,
as the former are usually associated with noise in real applications.
This property suggests that the weights should be nondecreasing,
\ie, $0\leq w_1\leq w_2\leq \ldots\leq w_r$.
However, we show that under this setting \rpotw is non-convex,
which is relevant from an optimization perspective.
A similar analysis was provided
by Chen et al. in \cite{chen2013reduced} for the adaptive nuclear norm;
we extend their results to the family of weighted Huber-based low-rank regularizers.
Interestingly, if \pot is selected as the hyperbola function,
the regularizer in \eref{e,reg_weight}
can be interpreted as a smooth version of the adaptive nuclear norm \cite{chen2013reduced}.
Analogously to the analysis we provide for \rpot,
we study the convexity and differentiability of \rpotc
and the Lipschitz continuity of its gradient.
We show that the proposed optimization framework can be extended
such that \rpotc can be used instead of \rpot.

The explicit gradient calculation of the cost function,
jointly with the step-size selection strategy provided in this work,
can be seen as the main steps of a smooth optimization framework
for inverse problems using low-rank models.
Importantly,
by enabling standard gradient-based iterative optimization algorithms,
it is possible to solve \eref{e,kost_local_h}
in the case of overlapping patches while also providing convergence guarantees.
Moreover, the same optimization algorithm can be used for different choices of \pot
when Huber's conditions are satisfied.
This offers users a convenient mechanism to explore different regularizers;
however, this flexibility might decrease computational speed
compared to algorithms that are designed for a particular non-smooth regularizer.
We show in the context of dynamic MRI reconstruction
that our proposed framework can 
converge faster and provide better image quality than
state-of-the-art algorithms (\eg, POGM)
that are tailored to commonly used non-smooth regularizers.

The main contributions of this work correspond to the following:
\begin{itemize}
    \item  We provide a smooth optimization framework to solve inverse problems
    involving either global or local low-rank models, 
    which are imposed through regularizers constructed using Huber-type smooth functions.
    \item We establish a theoretical framework based on singular value function theory
    to study relevant properties of the proposed regularizers 
    (\eg, convexity, differentiability, Lipschitz continuity of the gradient).
    \item We propose a step-size selection strategy 
    based on a MM approach 
    that monotonically decreases the line-search function. 
    For this purpose, we provide novel majorizers for the proposed regularizers 
    that exploit their Huber characteristics.
    \item We empirically show that,
    by using the proposed optimization framework, 
    standard smooth optimization algorithms 
    with convergence guarantees can be used  
    to solve inverse problems involving local low-rank models
    with overlapping patches. 
    Specifically, we show the application of the proposed smooth optimization framework
    in the context of dynamic MRI reconstruction.    
\end{itemize}

This paper is organized as follows.
Section \ref{sec,huber_reg} provides a detailed description
of the proposed Huber-based low-rank regularizers;
their convexity and differentiability properties are studied
by establishing connections with singular value functions.
Section \ref{sec,maj_reg} introduces novel majorizers
for the proposed regularizers
that are used to show the Lipschitz continuity of the regularizer gradient,
and to provide a quadratic majorizer for the line search function in later sections.
Section \ref{sec,weighted_reg} extends the proposed theoretical framework
to weighted Huber-based low-rank regularizers.
Section \ref{sec,opt} provides a smooth optimization framework
based on the proposed Huber-based low-rank regularizers
to minimize \eref{e,kost_global_h} or \eref{e,kost_local_h}
using standard iterative gradient-based optimization algorithms.
Section~\ref{sec,mri} illustrates the applicability of the proposed optimization framework
in the context of dynamic MRI reconstruction
using local low-rank models with overlapping patches.
Finally, Section \ref{sec,dis} provides discussion and conclusions.


\section{Huber-based low-rank regularizers}
\label{sec,huber_reg}

Let $\X\in\CMN$ be a rectangular matrix with singular values
$\sigma_1(\X) \geq \sigma_2(\X) \geq \ldots \geq \sigma_r(\X) \geq 0$
where $r = \min(M, N)$,
and let the vector-valued function $\vsig:\CMN \arrow \reals^r$
defined as
$\vsig(\X) \defequ [\sigma_1(\X), \ldots, \sigma_r(\X)]^T$.
Using this notation,
we write the proposed low-rank regularizer in \eref{e,reg_smooth} as:
\be
\rpot(\X) = (f_{\pot} \circ \vsig)(\X),
\ee{e,reg_comp}
where $\circ$ denotes function composition;
$f_{\pot}:\reals^r \arrow \reals$ is defined as
$f_{\pot}(\x) \defequ \sum_{k=1}^r \pot(x_k)$;
and $\pot:\reals \arrow \reals$
corresponds to a function satisfying the Huber conditions \cite[p. 184]{huber:81},
which we refer as the \textit{potential function}.
The extra notation introduced to write the regularizer expression in \eref{e,reg_comp}
might seem redundant,
but it facilitates the study of the properties of \rpot in later sections.
In fact, the expression in \eref{e,reg_comp}
reveals that \rpot corresponds to a singular value function
\cite{lewis1995convex, lewis2005nonsmooth}
that involves the potential function,
as indicated later in this section.
Notably, relevant properties for the optimization of the regularizer
highly depend on the analogous properties of the potential function.
For this reason we provide a detailed description of the latter in the following subsection. 
\begin{figure}[t] 
\centering 
\includegraphics[width=\textwidth]{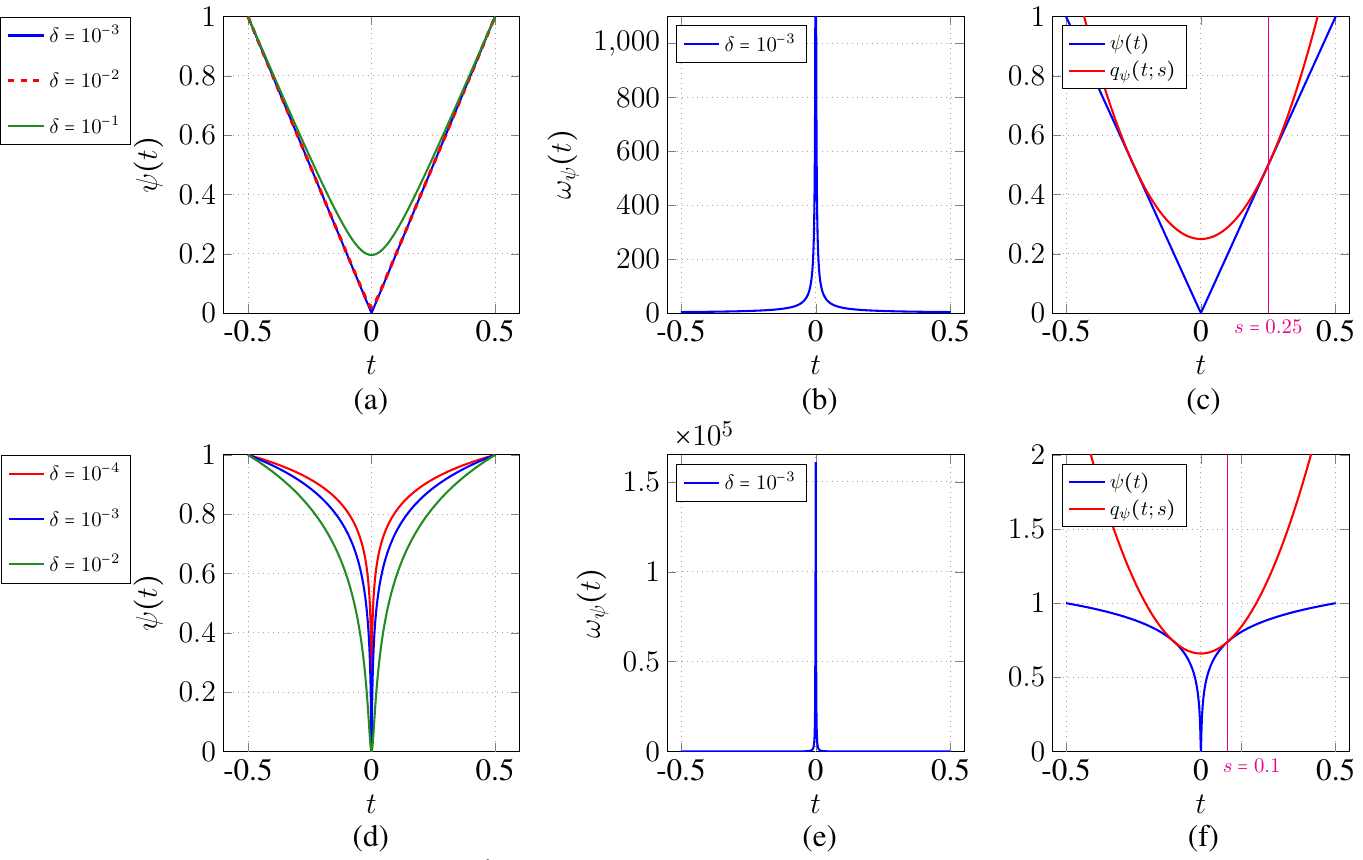} 
\caption{Examples of Huber potential functions.
(a) Hyperbola function (convex) as defined in \eref{e,pot_hyperbola} for three values of $\delta$.
Each curve has been normalized for illustration purposes.
The respective potential weighting function is shown in (b) for $\delta = 10^{-3}$.
For this case (c) shows the quadratic majorizer as defined in \eref{e,sur_pot}
for a particular value of $s$.
(d,e,f) show analogous results for the Cauchy function (non-convex)
as defined in \eref{e,pot_cauchy}.}
\label{fig:pot}
\end{figure}

\subsection{Huber potential functions}
\label{ss,huber_pot}
The potential function \pot in \eref{e,reg_smooth} and \eref{e,reg_comp}
satisfies the following Huber conditions \cite[p. 184]{huber:81}:
\begin{enumerate}[(i)]
\item
\pot is differentiable.
\item
\pot is even, \ie, $\pot(t) = \pot(-t), \forall t \in \reals$.
\item
$\wpot(t) \defequ \dpot(t)/t$,
where \dpot denotes the derivative of \pot,
is bounded, nonnegative and monotone nonincreasing for $t>0$.
This function is called the \textit{weighting function}.
\end{enumerate}
This family of functions,
collectively called Huber potential functions,
is extensive, and includes convex and non-convex functions
that have been widely used in inverse problems, 
e.g.,
\cite{holland:77:rru,charbonnier:97:dep,fessler:99:cgp}.
The assumption that \wpot is nonnegative
means that \dpot must be nonnegative for $t>0$,
which means that \pot must be nondecreasing for $t>0$.
There are examples of redescending potential functions
in the imaging literature \cite{chu:98:eps, yang:16:rlr}
(although perhaps not in terms of singular values).
We leave extensions to descending \pot for future work.

A representative example
for \pot is the hyperbola (\cf \fref{fig:pot}.a):
\be
\pot(t) = \del^2 \sqrt{1 + (t/\del)^2}
,\ee{e,pot_hyperbola}
which is approximately
$\del \abs{t}$
for $t \gg \del > 0$,
so \rpot can approximate
the nuclear norm
very well
by taking \del to be small (\cf \fref{fig:pot}.a).
Non-convex (but still smooth) choices for \pot
include the Cauchy function
(\cf \fref{fig:pot}.d):
\be
\pot(t) = \frac{1}{2}\delta^2 \log\left(1 + \left(\frac{t}{\delta}\right)^2\right).
\ee{e,pot_cauchy}

A notable property of a potential function \pot
is that it always has a quadratic majorizer
having curvature \wpot.
Specifically, for any $s\in \reals$, the function\footnote{Throughout the paper we use the notation $f(\cdot ~;~ \cdot)$
to indicate that the function $f$ depends on parameters
that are given after the semicolon symbol $;$ .}
\be
q_{\pot}(t; s) \defequ \pot(s) + \dpot(s)(t - s) + \frac{1}{2} \wpot(s) (t-s)^2
\ee{e,sur_pot}
satisfies:
\begin{eqnarray}
\pot(s) &=& q_{\pot}(s; s), \label{e,maj_a}\\
\pot(t) &\leq& q_{\pot}(t; s), \quad \forall t\in \reals. \label{e,maj_b}
\end{eqnarray}
Moreover,
$q_{\pot}$ is the quadratic majorizer for \pot
satisfying \eref{e,maj_a} and \eref{e,maj_b} with optimal (i.e., smallest) curvature
\cite[p.~186]{huber:81}.
This surrogate will be particularly useful when studying majorizers for \rpot
in Section \ref{sec,maj_reg}.
Interestingly, this surrogate also allows one to show that \dpot is Lipschitz continuous.
Given that \wpot is symmetric and monotonically nonincreasing for $t>0$, it follows that: 
\begin{align}
\pot(t) \leq q_{\pot}(t;s) & \leq \pot(s) + \dpot(s)(t-s)
+ \frac{1}{2}\left(\sup_{t\in \reals}\wpot(t)\right)(t-s)^2,
\quad \forall t, s \in \reals,\nonumber \\
& = \pot(s) + \dpot(s)(t-s) + \frac{1}{2}\wpot(0)(t-s)^2, \label{e,dpot_lips}
\end{align}
which is equivalent to \dpot being $\wpot(0)$-Lipschitz
\cite[Thm.~5.8]{beck:17:fom}.

Figure \ref{fig:pot} shows two examples of potential functions
with their respective weighting functions and quadratic majorizers.
Figure \ref{fig:pot}.a and \fref{fig:pot}.d illustrate
the (convex) hyperbola 
and the (non-convex) Cauchy potential functions,
respectively. 

The following subsection shows that many of the properties of the potential function $\pot$
(\eg, convexity, differentiability, Lipschitz continuity of the derivative)
are transferred to the regularizer \rpot
as a direct consequence of \rpot being a singular value function.
For this purpose, in the following derivations we consider $\CMN$ as a vector space
endowed with the (Frobenius)  
inner product
$\inprod{\cdot~}{\cdot}: \CMN \times \CMN \arrow \complex$
defined as: 
\be
\inprod{\A}{\B} = \trace{\A\B'},~\forall \A, \B \in \CMN,
\ee{e,inner_prod}
where $(\cdot)'$ denotes the Hermitian transpose operation.

\subsection{Properties of Huber-based low-rank regularizers}
\label{ss,reg_prop}

The regularizer proposed in \eref{e,reg_smooth} and \eref{e,reg_comp}
is contained in a class of functions known as unitarily invariant functions
or \textit{singular value functions}
\cite{lewis2005nonsmooth,lewis2005nonsmoothII,lewis1995convex,lewis1996convex,lewis2001twice},
which in general terms are functions
that depend only on the singular values of the matrix argument.
For completeness, we provide the formal definition in the following.

\begin{definition}[Unitarily invariant function or singular value function%
\cite{lewis1995convex,lewis2005nonsmooth,lewis2005nonsmoothII}]
A function $F:\CMN \arrow [-\infty, +\infty]$ is unitarily invariant
if $F(\U\X\V) = F(\X)$ for any $\X\in\CMN$ and any suitably sized unitary matrices \U and \V.
$F(\X)$ is also called a singular value function
as it depends only on the singular values of \X. 
\end{definition}

To facilitate the analysis of the proposed regularizers we also need the following definition.

\begin{definition}[Absolutely symmetric function \cite{lewis1995convex}]
A function $f:\reals^r \arrow [-\infty, +\infty]$
is absolutely symmetric
if $f(\x) = f(\check{\x})$
for any vector $\x\in\reals^r$,
where $\check{\x}\in\reals^r$ corresponds to the vector
formed by arranging the entries of the vector
$|\x|\defequ [|x_1|, \ldots, |x_r|]^T$ in nonincreasing order. 
\end{definition}

There is a one-to-one correspondence between unitarily invariant functions defined on $\CMN$
and absolutely symmetric functions defined on $\reals^r$ \cite{lewis1995convex}.
In fact, if $f:\reals^r \arrow [-\infty, +\infty]$ is absolutely symmetric,
then $f \circ \vsig$ is unitarily invariant.
Conversely, if $F:\CMN \arrow [-\infty, +\infty]$ is unitarily invariant,
then $f(\x) = F(\diag{\x})$ is absolutely symmetric,
where $\diag:\reals^r \arrow \CMN$
corresponds to a linear operator that creates a rectangular diagonal matrix
using the entries of its input.
Specifically, the entry in the $k$th row and $l$th column of $\diag{\x}\in \CMN$
is given by $[\diag{\x}]_{kl} = x_k$ if $k=l$ and $0$ otherwise.
Establishing the correspondence between singular value functions
and absolutely symmetric functions
is important, as many relevant properties of $F$
are directly related to the analogous properties
of the corresponding absolutely symmetric function $f$.
Particularly, convexity and differentiability were studied
by Lewis and Sendov
in \cite{lewis1995convex,lewis1996convex, lewis2005nonsmooth, lewis2005nonsmoothII}.
We summarize these results in the following proposition
as these properties are relevant to our optimization framework. 

\begin{proposition}[Convexity and differentiability of singular value functions \cite{lewis1995convex,lewis1996convex, lewis2005nonsmooth, lewis2005nonsmoothII}]
\label{th,cvx_dif_F}
Let $F = f \circ \vsig : \CMN \arrow [-\infty, +\infty]$ be a singular value function
with corresponding absolutely symmetric function
$f : \reals^r \arrow [-\infty, +\infty]$.
Then,

(i)
$F$ is convex if and only if $f$ is convex.

(ii)
If $f$ is convex,
then $F$ is differentiable at $\X \in \CMN$
if and only if $f$ is differentiable at $\vsig(\X)$.
In this case the gradient of $F$ is given by:
\be
\nabla F(\X) = \U\diag{\nabla f(\vsig(\X))}\V',
\ee{e,F_grad}
where \U and \V are suitably sized unitary matrices
such that $\X = \U\diag{\vsig(\X)}\V'$.

(iii)
If $f$ is non-convex and locally Lipschitz around $\vsig(\X)$,
then $F$ is differentiable at \X
if and only if $f$ is differentiable at $\vsig(\X)$.
The gradient of $F$ is given by \eref{e,F_grad}.
\end{proposition}

We can study the nuclear norm as an example of the previous theorem.
The nuclear norm is clearly unitarily invariant,
and its corresponding absolutely symmetric function is given by
$f_*{(\x)} \defequ \sum_{k=1}^r|x_k|$,
\ie, $\mnorm{\X}_* = (f_* \circ \vsig)(\X)$.
The function $f_*$ also corresponds 
to a symmetric gauge function \cite{mirsky:60:sgf}. 
It can be shown that absolutely symmetric functions
correspond to symmetric gauge functions 
when, in addition,
they are positively homogeneous \cite{lewis:95:tca}.
Given that $f_*$ is convex and non-differentiable
at points where some of the entries are equal to zero,
then, according to \tref{th,cvx_dif_F},
$\mnorm{\cdot}_*$ is convex and non-differentiable
at matrices where some of the singular values are equal to zero. 

In our setting,
it is straightforward to show that the function
\fpot in \eref{e,reg_comp} is absolutely symmetric
and therefore the corresponding regularizer \rpot is a singular value function
(\ie, unitarily invariant).
Then, we can establish conditions for the convexity and differentiability of \rpot
by analyzing \fpot and using Proposition \ref{th,cvx_dif_F}.
In our context,
these conditions are directly related to the potential function \pot
as stated in the next theorem.
Appendix~\ref{s,ap_proof_prop_dif}
gives the straightforward proof.

\begin{theorem}[Convexity and differentiability of Huber-based low-rank regularizers]
\label{th,cvx_dif_R}
Let \rpot be a Huber-based low-rank regularizer defined
as in \eref{e,reg_smooth} and \eref{e,reg_comp}.
Then, 

(i)
\rpot is convex if and only if \pot is convex.

(ii)
\rpot is differentiable at $\X \in \CMN$ and its gradient is given by:
\be
\nabla \rpot(\X) = \U \diag{\dpotdot{\vsig(\X)}} \V',
\ee{e,R_grad}
where \U and \V are suitably sized unitary matrices
such that $\X = \U \diag{\vsig(\X)} \V'$.
\end{theorem}
In \eqref{e,R_grad},
$\dpotdot{\vsig(\X)}$
denotes the vector formed by applying the function $\dpot$ element-wise to the vector $\vsig(\X)$,
\ie,
$\dpotdot{\vsig(\X)} = [\dpot(\sigma_1(\X)), \ldots, \dpot(\sigma_r(\X))]^T$.
This notation arises in the Julia language
\cite{bezanson:17:jaf}
and we use it throughout.

\begin{remark}
The differentiability of \rpot is independent of its convexity,
so we do not have two cases as Proposition \ref{th,cvx_dif_F} might suggest.
This is because \fpot is always locally Lipschitz
as $\nabla \fpot$ is Lipschitz continuous,
which happens even when \pot is non-convex.
This follows from the fact that \dpot is Lipschitz continuous (\cf, \eref{e,dpot_lips}).
Appendix~\ref{s,ap_proof_prop_dif}
provides more details
in the proof of \tref{th,cvx_dif_R}.
\end{remark}

Another important property to study is the Lipschitz continuity of $\nabla\rpot(\X)$.
Specifically, we want to determine conditions under which $\nabla \rpot$ is Lipschitz continuous.
By following the same principles in \tref{th,cvx_dif_R},
the next proposition shows that these conditions are directly related
to the Lipschitz properties of \dpot.
Appendix \ref{th,proof_smooth_R} gives a proof for this proposition
based on the results in Section \ref{sec,maj_reg}.
\begin{proposition}[Lipschitz continuity of the gradient of Huber-based low-rank regularizers]
\label{th,smooth_R}
Let \rpot be a regularizer defined as in \eref{e,reg_smooth} and \eref{e,reg_comp}.
Then, $\nabla\rpot$ is $L$-Lipschitz continuous where $L = \wpot(0)$,
i.e.,
\be
\mnormfrob{\nabla\rpot(\X) - \nabla\rpot(\Y)}
\leq \wpot(0)\mnormfrob{\X - \Y}, ~\forall \X, \Y \in \CMN.
\ee{e,grad_lips}
\end{proposition}
The properties shown in \tref{th,cvx_dif_R} and Proposition \ref{th,smooth_R}
suffice to enable standard iterative gradient-based optimization methods
for solving \eref{e,kost_global_h} and \eref{e,kost_local_h} as shown in Section \ref{sec,opt};
however, the fast convergence of these optimization methods
highly depends on the adopted step-size selection rule.
The next section proposes a new class of majorizers for Huber-based low-rank regularizers.
These majorizers are then used to propose efficient step-size selection strategies
that enable the fast convergence of iterative gradient-based optimization methods
when using Huber-based low-rank regularizers. 

\section{Majorizers for Huber-based low-rank regularizers and their associated line search functions}
\label{sec,maj_reg}
The previous section showed that many relevant properties of the potential function \pot
are transferred to the Huber-based low-rank regularizer \rpot.
Following this principle,
and given that quadratic majorizers can be defined for \pot as shown in \eref{e,sur_pot},
this section derives novel majorizers for \rpot.
Inspired by the work of Lewis and Sendov in \cite{lewis2002quadratic},
that derives a quadratic expansion
to approximate spectral functions defined on real symmetric matrices,
here we provide a general expression for majorizers for Huber-based low-rank regularizers.
Our results are directly stated in the more general case
where Huber-based low-rank regularizers are defined on complex rectangular matrices.

A majorizer for \rpot at $\S\in\CMN$, denoted by $J(\cdot~; \S)$, 
corresponds to a matrix-valued function that satisfies:
\begin{eqnarray}
\rpot(\S) &=& J(\S; \S), \label{e,maj_a_reg}\\
\rpot(\X) &\leq& J(\X; \S), \quad \forall \X\in \CMN. \label{e,maj_b_reg}
\end{eqnarray}
We show below that \eref{e,maj_a_reg} and \eref{e,maj_b_reg}
are satisfied for functions with the form:
\be
Q(\X; \S, \gpot) = \rpot(\S) + \rinprod{\nabla \rpot(\S)}{\X -\S}
+ \frac{1}{2}(\vone_M'(\gpot(\vsig(\S)) \odot |\U'(\X -\S)\V|.^{\wedge}2) \vone_N),
\ee{e,maj_reg_gen}
where $\vone_K$ denotes a vector of length $K$ with each entry equal to one;
$\odot$ denotes entrywise (Hadamard) product;
\U and \V are suitably sized unitary matrices such that $\S = \U \diag{\vsig(\S)} \V'$;
and $\gpot : \reals^r \arrow \CMN$ is an operator that depends on \pot
that constructs a matrix out of the entries of its input.
Finally, the notation $|\A|.^{\wedge}2$ for $\A\in\CMN$
corresponds to a matrix in \RMN whose entries are the entries of \A
after taking the absolute value and square operations entry-wise.
If $\B = |\A|.^{\wedge}2$, then $B_{kl} = |A_{kl}|^2$.
The derivation for the expression in \eref{e,maj_reg_gen}
was inspired on the results found in \cite{lewis2002quadratic},
where the authors derive a quadratic expansion to approximate spectral functions. 
In their case, the third term in the expansion
involves an entrywise multiplication using specific factors
that depend on first derivatives.
For our case, we made the observation that 
these factors could be modified
such that $Q(\X; \S, \gpot)$ could be a majorizer for \rpot.
These factors compose the entries of the matrix given by the output of $\gpot$.
In the following, we show that different choices for $\gpot$ 
provide different majorizers for \rpot,
this is, $Q(\X; \S, \gpot)$
satisfy \eref{e,maj_a_reg} and \eref{e,maj_b_reg}.
Particularly, we show two options for the operator $\gpot$.
The first operator is defined in terms of the weighting function \wpot
and is denoted by \gpotR.
If $\vs \in \reals^r$ 
and $M<N$ (wide case), \gpotR is defined as: 
\begin{eqnarray}
\gpotR(\vs) &\defequ& (\wpotdot{\vs}) \, \vone_N',
\label{e,gpot_R_wide}
\end{eqnarray}
and when $M>N$ (tall case):
\begin{eqnarray}
\gpotR(\vs) &\defequ& \vone_M \, (\wpotdot{\vs})'
\label{e,gpot_R_tall}.
\end{eqnarray}%
The second operator is defined in terms of the Lipschitz constant
of the gradient of \rpot (\cf, Proposition \ref{th,smooth_R}), 
and is denoted by \gpotL.
For both the wide and tall cases this operator is defined as:
\begin{eqnarray}
\gpotL(\vs) &\defequ& \wpot(0) \vone_M \vone_N'.
\label{e,gpot_L_wide}
\end{eqnarray}

The following theorem states that
$Q(\cdot~; \S, \gpotR)$ and $Q(\cdot~; \S, \gpotL)$
are majorizers for \rpot at $\S \in \CMN$,
and it also provides equivalent expressions for both functions.
Appendix \ref{th,proof_maj_Huber} provides its proof.

\begin{theorem}[Majorizers for Huber-based low-rank regularizers]
\label{th,maj_Huber}
Let $\S\in \CMN$
and $\U, \V$ suitably sized unitary matrices such that $\S = \U \diag{\vsig(\S)} \V'$.
Then $Q(\cdot~; \S, \gpotR)$ and $Q(\cdot~; \S, \gpotL)$
are both majorizers for \rpot at $\S \in \CMN$,
\ie, both satisfy \eref{e,maj_a_reg} and \eref{e,maj_b_reg}.
If $M<N$ (wide case), then 
\begin{align}
Q(\X; \S, \gpotR) &= \rpot(\S) + \rinprod{\nabla \rpot(\S)}{\X -\S}
 + \frac{1}{2}\mnormfrob{\W_{\pot, \S}' \, (\X -\S)}^2,
\label{e,maj_reg_gen_R_frob_w}
\end{align}
where the matrix $\W_{\pot, \S}$ is defined as%
\be
\W_{\pot, \S} \defequ \U \diag{\sqrt{\wpotdot{\vsig(\vs)}}} \V \in\CMN,
\ee{e,w_mat}
where the square-root is applied element-wise.
If $M>N$ (tall case), then 
\begin{align}
Q(\X; \S, \gpotR) &= \rpot(\S) + \rinprod{\nabla \rpot(\S)}{\X -\S}
 + \frac{1}{2}\mnormfrob{(\X -\S) \, \T_{\pot, \S}'}^2,
\label{e,maj_reg_gen_R_frob_t}
\end{align}
where the matrix $\T_{\pot, \S}$ is defined as:
\be
\T_{\pot, \S} \defequ \U' \diag{\sqrt{\wpotdot{\vsig(\vs)}}} \V' \in\CMN.
\ee{e,t_mat}
For the wide and tall cases we have that:
\be
Q(\X; \S, \gpotL) = \rpot(\S) + \rinprod{\nabla \rpot(\S)}{\X -\S}
 + \frac{1}{2}\wpot(0)\mnormfrob{\X -\S}^2.
\ee{e,maj_reg_gen_L_frob_w}
\end{theorem}

In the following we use \tref{th,maj_Huber}
to derive quadratic majorizers for the 1D line search function
associated with the proposed regularizer.
The construction of these quadratic majorizers is one of our main results
because it provides a simple
step-size selection strategy
that ensures monotonic descent
in the optimization framework proposed in Section \ref{sec,opt}. 

Using \tref{th,maj_Huber},
we show in the following that quadratic majorizers can be constructed
for the regularizer 1D line search function $\hpot: \reals \arrow \reals$ defined as:
\be
\hpot(\alf; \X, \bDel) \defequ \rpot(\X + \alf \bDel),
\ee{e,hpot}
where $\alf \in \reals$ and $\X, \bDel \in \CMN$ are two arbitrary (fixed) matrices.
The following theorem provides a general form for these quadratic majorizers.
Appendix \ref{th,hpot_proof} provides
its proof using \tref{th,maj_Huber}.

\begin{theorem}\label{th,hpot}
Let $\gpot:\reals^r \arrow \CMN$ be an operator such that
$Q(\cdot ~; \X + \alfs\bDel, \gpot)$, defined as in \eref{e,maj_reg_gen},
is a majorizer for \rpot at $\X + \alfs\bDel$
where $\alfs\in \reals$.
Define the quadratic function
\be
g_{\hpot}(\alf ; \alfs, \X, \bDel, \gpot) \defequ
c_{\hpot}^{(0)}(\alfs, \X, \bDel)
+ c_{\hpot}^{(1)}(\alfs, \X, \bDel)(\alf-\alfs)
+ \frac{1}{2}c_{\hpot}^{(2)}(\alfs, \X, \bDel, \gpot) (\alf - \alfs)^2
\ee{e,hpot_maj}
with coefficients given by:
\begin{align}
c_{\hpot}^{(0)}(\alfs, \X, \bDel) & = \rpot(\X + \alfs\bDel) = \hpot(\alfs; \X, \bDel), \\
c_{\hpot}^{(1)} (\alfs, \X, \bDel)& = \rinprod{\nabla \rpot(\X + \alfs\bDel)}{\bDel}, \\
c_{\hpot}^{(2)}(\alfs, \X, \bDel, \gpot) & =
\vone_M'(\gpot(\vsig(\X + \alfs\bDel)) \odot |\U'\bDel\V|.^{\wedge}2) \vone_N,
\end{align}
where $\U, \V$ are suitable sized unitary matrices such that
$\X + \alfs\bDel = \U\diag{\vsig(\X + \alfs\bDel)}\V'$.
Then, $g_{\hpot}(\alf ; \alfs, \X, \bDel, \gpot)$
is a quadratic majorizer for
$h_{\pot}(\alf; \X, \bDel)$ at \alfs.
\end{theorem}

Combining the results of \tref{th,maj_Huber} and \tref{th,hpot}
yields the following corollary that 
can guide the construction of quadratic majorizers for \hpot in practice.

\begin{corollary}\label{cr,maj}
$g_{\hpot}(\alf ; \alfs, \X, \bDel, \gpotR)$ and
$g_{\hpot}(\alf ; \alfs, \X, \bDel, \gpotL)$
are both quadratic majorizers for $h_{\pot}(\alf; \X, \bDel)$ at $\alfs$,
and
$$
g_{\hpot}(\alf ~ ; \alfs, \X, \bDel, \gpotR) \leq g_{\hpot}(\alf ~ ; \alfs, \X, \bDel, \gpotL),
~ \forall \alpha \in \reals.
$$
\end{corollary}
Appendix \ref{cr,maj_proof} provides the proof of this corollary.

Before showing how to use the previous results in the proposed optimization framework
in Section \ref{sec,opt},
the following section shows that all these previous results
extend to the proposed weighted Huber-based low-rank regularizers defined in \eref{e,reg_weight}.

\section{Weighted Huber-based low-rank regularizers}
\label{sec,weighted_reg}

The regularizer $\rpotc$ defined in \eref{e,reg_weight} is unitarily invariant
because it can be written as:
\be
\rpotc(\X) = (\fpotc \circ \vsig)(\X),
\ee{e,reg_comp_weight}
where $\fpotc:\reals^r \arrow \reals$
is defined as the absolutely symmetric function
\be
\fpotc(\x) \defequ \sumno_{k=1}^r w_k \pot([\check{\x}]_k).
\ee{e,fpotc}
Then, according to Proposition \ref{th,cvx_dif_F},
we can assess the convexity and differentiability of \rpotc by studying \fpotc. 
To simplify the following analysis,
we assume that \pot is not a constant 
and that \w is nonzero.
Both of these assumptions are reasonable in practice,
as a constant potential function or having all weights equal to zero
would not properly promote low-rank characteristics when constructing \rpotw.
Under these assumptions we begin studying the convexity of \fpotw,
which depends on the weights \w and the potential function \pot
as shown in the following proposition.

\begin{proposition}\label{prop,fpotc_cvx}
\fpotc is convex if and only if the Huber potential \pot is convex and
$w_1\geq w_2\geq \ldots \geq w_r\geq 0$.
\end{proposition}
Appendix \ref{ap,proof_fpot_cvx} provides the proof.
The following proposition provides conditions for the differentiability of $\fpotc$
at vectors that are relevant in our analysis.
Interestingly, these conditions do not depend on the vector $\w$.

\begin{proposition}\label{prop,fpotc_dif}
Let $\x \in \reals^r$ be a vector whose entries are unique, nonnegative,
and arranged in decreasing order,
\ie, $\x = \check{\x}$ and $x_k>x_{k+1}, ~\forall k\in\{1\ldots,r-1\}$. 
Then, \fpotc is locally Lipschitz and differentiable at \x with gradient given by:
\be
\nabla \fpotc(\x) = \w \odot \dpotdot{\x}.
\ee{e,fpotc_grad}
\end{proposition}
Appendix \ref{ap,proof_fpot_dif} provides the proof.

The previous results allow us to establish conditions
for the convexity and differentiability of \rpotc.
We summarize these results in the following theorem,
which is provided without proof
as it is a direct application of Proposition \ref{th,cvx_dif_F}
using Proposition \ref{prop,fpotc_cvx} and Proposition \ref{prop,fpotc_dif}.
\begin{theorem}[Convexity and differentiability of weighted Huber-based low-rank regularizers]
\label{th,wreg_cvx_dif}

(i) \rpotw is convex if and only if \pot is convex and $w_1\geq w_2\geq \ldots \geq w_r\geq 0$. 

(ii) If $\X\in\CMN$ has no repeated singular values, \ie,
$\sigma_1(\X) > \sigma_2(\X) > \ldots > \sigma_r(\X)\geq 0$,
then \rpotc is differentiable at \X and
\be
\nabla \rpotc(\X) = \U\diag{\w \odot \dpotdot{\vsig(\X)}}\V',
\ee{e,R_grad_weighted}
where \U and \V are suitably sized unitary matrices such that $\X = \U \diag{\vsig(\X)} \V'$. 
\end{theorem}

\begin{remark}
In applications we might like low-rank regularizers
to penalize small singular values more than large singular values,
as the former are usually associated with noise.
This preference would suggest using weights \w
whose components are in nondecreasing order,
\ie, $0\leq w_1\leq w_2\leq \ldots\leq w_r$.
Under this setting \tref{th,wreg_cvx_dif} shows that \rpotw would be non-convex,
which can complicate optimization.
We show empirically in the following sections
that in this non-convex setting \rpotc can outperform its non-weighted version \rpot
when a good initialization is provided.
\end{remark}

\begin{remark}
\tref{th,wreg_cvx_dif} provides a closed-form expression for $\nabla \rpotc(\X)$,
whose existence is based on the assumption
that \X does not have repeated singular values.
This allows us to use Proposition \ref{th,cvx_dif_F} with Proposition \ref{prop,fpotc_dif};
in the case of repeated singular values the function \fpotw 
would not be necessarily differentiable at $\vsig(\X)$.
This condition might seem limiting as low-rank matrices present repeated singular values.
Methods based on the nuclear norm
involve thresholding operations
that lead to ``exactly'' zero singular values.
In contrast,
our smooth approach will shrink singular values towards zero,
but never thresholds them exactly to zero,
so repeated zero singular values are extremely unlikely.
In addition, if the gradient $\nabla \rpotc(\X)$ (as given in \eref{e,R_grad_weighted})
is calculated by computing a singular value decomposition (SVD) of \X
using a computational solver, 
as is usually done in real applications, 
it is highly unlikely to obtain repeated singular values
due to the presence of noise and numerical precision errors.
Therefore, hereafter we assume that the sufficient condition of
\tref{th,wreg_cvx_dif} corresponding to having no repeated singular values
is satisfied when calculating the gradient of \rpotc. 
\end{remark}

Under the setting of nondecreasing weights
it is also possible to construct majorizers for \rpotc
that are analogous to the ones provided for \rpot in \eref{e,maj_reg_gen}.
If $\w \in \reals^r$ such that $0\leq w_1\leq w_2\leq \ldots\leq w_r$,
then these majorizers have the form:
\begin{align}
\qpotw(\X; \S, \gpotw) &= \rpotw(\S) + \rinprod{\nabla \rpotw(\S)}{\X -\S}
\nonumber \\
& + \frac{1}{2}(\vone_M'(\gpotw(\vsig(\S)) \odot |\U'(\X -\S)\V|.^{\wedge}2) \vone_N),
\label{e,maj_reg_gen_weighted}
\end{align}
where $\U, \V$ are suitably sized unitary matrices such that $\S = \U\diag{\vsig(\S)}\V'$
and $\gpotw: \reals^r \arrow \CMN$ is an operator that depends on \pot and \w.
Appendix \ref{s,ap_weight} shows that two operators
denoted by \gpotwR and \gpotwL
provide valid majorizers for \rpotw.
Specifically, we state an extended version of \tref{th,maj_Huber}.
Using this theorem we can show that the gradient of \rpotw is Lipschitz continuous. 

\begin{proposition}[Lipschitz continuity
of the gradient of weighted Huber-based low-rank regularizers]
\label{th,smooth_R_weighted}
For
nonnegative
and nondecreasing
$\w \in \reals^r$,
$\nabla\rpotw$ is $L$-Lipschitz continuous where
$L = \norminf{\w} \wpot(0)$,
i.e.:
\be
\mnormfrob{\nabla\rpotw(\X) - \nabla\rpotw(\Y)} \leq
\norminf{\w} \wpot(0) \mnormfrob{\X - \Y},
~\forall \X, \Y \in \CMN.
\ee{e,grad_lips_weighted}
\end{proposition}
Appendix \ref{s,ap_proof_smooth_R_weighted} provides the proof.

Using the previous results it is also possible to construct quadratic majorizers
for the 1D function
$ \hpotw(\alf; \X, \bDel) \defequ \rpotw(\X + \alf \bDel)$.
The procedure is analogous to the one described in \tref{th,hpot}
and is not provided for the sake of space.

One important case is when the vector of weights is defined
such that the first $K$ entries are equal to zero
and the rest are equal to one.
Denoting this vector by $\w_K$, it follows that: 
\be
[\w_K]_k \defequ \begin{cases}
0 & \text{if $k \leq K$}, \\
1 & \text{otherwise,}
\end{cases}
\ee{e,weights_tail}
where $K\in\{0,\ldots, r-1\}$.
If $K$ is a good estimate of the true rank of \X,
then \rpotwk will not penalize the first $K$ singular values, as desired.
It would only penalize the tail singular values.
We call this version of the weighted regularizer
the \textit{tail Huber-based low-rank regularizer}:
\be
\rpotk(\X) \defequ R_{\pot, \w_K}(\X) = \sum_{k=K+1}^r\pot(\sig_k(\X)).
\ee{e,reg_tail}
This regularizer is non-convex given that the entries of $\w_K$ are in nondecreasing order
(see \tref{th,wreg_cvx_dif} part (i)).
Nevertheless, we can calculate its gradient using \tref{th,wreg_cvx_dif} part (ii)
and also construct majorizers for it using \tref{th,maj_Huber_weighted}.
Therefore, it is possible to construct quadratic majorizers
for the associated 1D line search function using the procedure in \tref{th,hpot}.

At this point we have derived a closed-form expression
for the gradients of the proposed regularizers,
and also provided quadratic majorizers for the associated 1D line search functions.
Section \ref{sec,opt} leverages these results
to solve optimization problems \eref{e,kost_global_h} and \eref{e,kost_local_h}
using standard iterative gradient-based optimization methods.

\section{Smooth optimization framework using Huber-based low-rank regularizers}
\label{sec,opt}

Instead of promoting low-rank models using the (non-smooth) nuclear norm
in \eref{e,kost} and \eref{e,kost_local},
we propose to use smooth Huber-based low-rank regularizers%
\footnote{The following analysis can be easily extended
to weighted Huber-based low-rank regularizers.
We omit those derivations for the sake of notation simplicity.},
\ie, \rpot as defined in \eref{e,reg_smooth},
to minimize \eref{e,kost_global_h} and \eref{e,kost_local_h}. 

Since \eref{e,kost_local_h} corresponds to a more general model
than%
\footnote{We can recover \eref{e,kost_global_h} from \eref{e,kost_local_h}
by selecting $\jcal{P}_{\vp}$ and $\jcal{S}_{\vs}$ as the identity operator
and scaling \reg according to the cardinality of $\Lambda$ and $\Gamma$.}
\eref{e,kost_global_h},
we focus on the local low-rank optimization problem.
To use iterative gradient-based optimization algorithms to minimize $\kost_{\text{Local}}$,
we start by calculating its gradient by using \tref{th,cvx_dif_R}.
This gradient is simply given by: 
\be
\nabla \kost_{\text{Local}}(\X) = \nabla f(\X)
+ \reg \sum_{\vs \in \Lambda}\sum_{\vp \in \Gamma}
\jcal{S}_{\vs}^*(\jcal{P}_{\vp}^*(\nabla \rpot(\jcal{P}_{\vp}(\jcal{S}_{\vs}(\X))))),
\ee{e,kost_llr_grad}
where $\nabla\rpot(\jcal{P}_{\vp}(\jcal{S}_{\vs}(\X)))$ is calculated using \eref{e,R_grad},
and $\jcal{S}_{\vs}^*$ and $\jcal{P}_{\vp}^* $
correspond to the the adjoint operators of $\jcal{S}_{\vs}$ and $\jcal{P}_{\vp}$, respectively.
This expression enables the minimization of $\kost_{\text{Local}}$
using gradient-based optimization algorithms like NCG or quasi-Newton methods;
however, the fast convergence of these iterative algorithms
highly depends on a step-size selection strategy
that guarantees a monotonic decrease of $\kost_{\text{Local}}$ in each iteration.
The ideal step size could be determined
by minimizing the line search function:
\be
h(\alf; \X, \bDel) \defequ \kost_{\text{Local}}(\X + \alf \bDel),
\ee{e,line-search}
where $\X, \bDel \in \CMN$ are two arbitrary (fixed) matrices.

For notation simplicity we define the matrices
$\X^{(\vp, \vs)} \defequ \jcal{P}_{\vp}(\jcal{S}_{\vs}(\X))$
and $ \bDel^{(\vp, \vs)} \defequ \jcal{P}_{\vp}(\jcal{S}_{\vs}(\bDel))$
where $(\vp, \vs) \in \Gamma \times \Lambda$,
and we also define the 1D functions
\begin{align}
h_f(\alf; \X, \bDel) & \defequ f(\X + \alf \bDel), \label{e,h_f}\\
h_{\pot, \text{Local}}(\alf; \X, \bDel) & \defequ \sum_{\vs \in \Lambda}\sum_{\vp \in \Gamma}
h_{\pot}(\alf; \X^{(\vp, \vs)}, \bDel^{(\vp, \vs)}),
\label{e,h_local}
\end{align}
where $\hpot$ was defined in \eref{e,hpot}.
Then, we can express the line search function $h$ as follows:
\begin{align}
h(\alf~; \X, \bDel)
& = \kost_{\text{Local}}(\X + \alf \bDel) \nonumber \\
& = f(\X + \alf \bDel) + \reg \sum_{\vs \in \Lambda}\sum_{\vp \in \Gamma}
\rpot(\jcal{P}_{\vp}(\jcal{S}_{\vs}(\X + \alf \bDel))) \nonumber \\
& = f(\X + \alf \bDel) + \reg \sum_{\vs \in \Lambda}\sum_{\vp \in \Gamma}
\rpot(\jcal{P}_{\vp}(\jcal{S}_{\vs}(\X)) + \alf \jcal{P}_{\vp}(\jcal{S}_{\vs}(\bDel))) \nonumber \\
& = h_f(\alpha; \X, \bDel) + \reg \sum_{\vs \in \Lambda}\sum_{\vp \in \Gamma}
h_{\pot}(\alf; \X^{(\vp, \vs)}, \bDel^{(\vp, \vs)}) \nonumber \\
& = h_f(\alpha; \X, \bDel) + \reg h_{\pot, \text{Local}}(\alf; \X, \bDel),
\label{e,line_freg} 
\end{align}
where the third equality follows from the linearity of $\jcal{P}_{\vp}$ and $\jcal{S}_{\vs}$.
Thus, \eref{e,line_freg} reveals that
$h(\alf~; \X, \bDel)$ can be minimized
by addressing the minimization of the functions
$h_f(\alpha; \X, \bDel)$ and $h_{\pot, \text{Local}}(\alf; \X, \bDel)$.
Our strategy for this purpose is based on using quadratic majorizers for each of these 1D functions.
In many inverse problems (\eg, dynamic MRI reconstruction),
$f$ has a quadratic structure and therefore it is straightforward
to derive a quadratic majorizer for $h_f(\alpha; \X, \bDel)$.
Then, for the following derivations we assume that there exists a quadratic function defined as:
\be
g_{f}(\alf~; \alfs, \X, \bDel) \defequ c_{f}^{(0)}(\alfs, \X, \bDel)
+ c_{f}^{(1)}(\alfs, \X, \bDel)(\alf-\alfs)
+ \frac{1}{2}c_{f}^{(2)}(\alfs, \X, \bDel) (\alf - \alfs)^2,
\ee{e,hf_maj}
that is a majorizer for $h_f(\alf; \X, \bDel)$ at $\alfs \in \reals$,
where $\{ c_{f}^{(l)}(\alfs, \X, \bDel)\}_{l=0}^2$ are three scalar coefficients
that depend on $\alfs, \X$ and $\bDel$.
We use \tref{th,hpot} and Corollary \ref{cr,maj}
to construct a quadratic majorizer for $h_{\pot, \text{Local}}(\alf; \X, \bDel)$.
The following theorem provides this construction.
\begin{theorem}\label{th,h_local_maj}
Let $\gpot$ be an operator%
\footnote{The input and output dimensions of $\gpot$ are different
in \eref{e,hpot_maj} and \eref{e,h_maj}.
We have adopted the same notation in both cases for simplicity,
as the dimensions should be clear from context.}
such that $Q(\cdot ~; \X^{(\vp, \vs)} + \alfs\bDel^{(\vp, \vs)}, \gpot)$,
defined as in \eref{e,maj_reg_gen},
is a majorizer for \rpot at
$\X^{(\vp, \vs)} + \alfs\bDel^{(\vp, \vs)}, ~ \forall (\vp, \vs) \in \Gamma \times \Lambda$.
Then, the quadratic function 
\begin{align}
g_{h_{\pot, \mathrm{Local}}}(\alf~; \alfs, \X, \bDel, \gpot)
&\defequ c_{h_{\pot, \mathrm{Local}}}^{(0)}(\alfs, \X, \bDel)
+ c_{h_{\pot, \mathrm{Local}}}^{(1)}(\alfs, \X, \bDel) \, (\alf-\alfs)
\nonumber \\
& \quad
+ \frac{1}{2}c_{h_{\pot, \mathrm{Local}}}^{(2)}(\alfs, \X, \bDel, \gpot) \, (\alf - \alfs)^2,
\label{e,h_maj}
\end{align}
with coefficients given by:
\begin{align}
c_{h_{\pot, \mathrm{Local}}}^{(0)}(\alfs, \X, \bDel) &=
\sum_{\vs \in \Lambda}\sum_{\vp \in \Gamma}
c_{\hpot}^{(0)}(\alfs, \X^{(\vp, \vs)},\bDel^{(\vp, \vs)}),
\nonumber \\
c_{h_{\pot, \mathrm{Local}}}^{(1)}(\alfs, \X, \bDel) &=
\sum_{\vs \in \Lambda}\sum_{\vp \in \Gamma}
c_{\hpot}^{(1)}(\alfs, \X^{(\vp, \vs)},\bDel^{(\vp, \vs)}),
\label{e,c1_pot_local} \\
c_{h_{\pot, \mathrm{Local}}}^{(2)}(\alfs, \X, \bDel, \gpot) &=
\sum_{\vs \in \Lambda}\sum_{\vp \in \Gamma}
c_{\hpot}^{(2)}(\alfs, \X^{(\vp, \vs)}, \bDel^{(\vp, \vs)}, \gpot),
\label{e,c2_pot_local}
\end{align}
is a majorizer for $h_{\pot, \mathrm{Local}}(\alf~; \X, \bDel)$ at \alfs.
\end{theorem}

\begin{proof}
It follows directly from using \eref{e,hpot_maj}
to provide a quadratic majorizer for each term
in the sum on the right-hand-size of \eref{e,h_local}. 
\end{proof}
Given that we have constructed quadratic majorizers for $h_f$ and $h_{\pot, \text{Local}}$,
the following corollary of \tref{th,h_local_maj}
derives a quadratic majorizer for the line search function $h$.
\begin{corollary}\label{cr,h_maj}
Let $\gpot$ be an operator defined as in \tref{th,h_local_maj}.
Then, the quadratic function
\begin{align}
g_{h}(\alf~; \alfs, \X, \bDel, \gpot) & \defequ g_{h_f}(\alf~; \alfs, \X, \bDel)
+ \reg g_{h_{\pot, \mathrm{Local}}}(\alf~; \alfs, \X, \bDel, \gpot)
\label{e,g_maj}\\
& = c_{h}^{(0)}(\alfs, \X, \bDel) + c_{h}^{(1)}(\alfs, \X, \bDel)(\alf-\alfs)
\nonumber \\
&\quad + \frac{1}{2}c_{h}^{(2)}(\alfs, \X, \bDel, \gpot) (\alf - \alfs)^2,
\end{align}
with coefficients given by:
\begin{align}
c_{h}^{(0)}(\alfs, \X, \bDel) &=
c_{f}^{(0)}(\alfs, \X, \bDel) + \reg c_{h_{\pot, \mathrm{Local}}}^{(0)}(\alfs, \X, \bDel),
\nonumber \\
c_{h}^{(1)}(\alfs, \X, \bDel) &= 
c_{f}^{(1)}(\alfs, \X, \bDel) + \reg c_{h_{\pot, \mathrm{Local}}}^{(1)}(\alfs, \X, \bDel),
\label{e,h_c1}\\
c_{h}^{(2)}(\alfs, \X, \bDel, \gpot) &= 
c_{f}^{(2)}(\alfs, \X, \bDel) + \reg
c_{h_{\pot, \mathrm{Local}}}^{(2)}(\alfs, \X, \bDel, \gpot),
\label{e,h_c2}
\end{align}
is a majorizer for $h(\alf~; \X, \bDel)$ at $\alfs$.
\end{corollary}
\begin{remark}
The existence of a majorizer for the line search function $h$
given by Corollary \ref{cr,h_maj}
enables the minimization of $h$
using a simple iterative majorize-minimize approach (MM)
\cite[\S~IV]{fessler:99:cgp}.
Given an initial guess $\alf_0$,
it is straightforward to show that the iterative step
\begin{align}
\alf _{k+1} \defequ \argmin{\alf \in \reals} g_{h}(\alf ~; \alf_k, \X, \bDel, \gpot)
& = \alf_{k} - \frac{c_{h}^{(1)}(\alf_k, \X, \bDel)}{c_{h}^{(2)}(\alf_k, \X, \bDel, \gpot)}
\label{e,ss_iter}
\end{align}
decreases $h(\alf~; \X, \bDel)$ monotonically,
i.e., $h(\alf_{k+1}~; \X, \bDel)\leq h(\alf_k~; \X, \bDel)$.
Let $n_{\alf}$ denote the number of iterations
of \eref{e,ss_iter} that we use to descend $h(\alf~; \X, \bDel)$.
\end{remark}

To apply the MM step-size selection method in \eref{e,ss_iter},
one must calculate $c_{h_{\pot, \text{Local}}}^{(1)}$ and $c_{h_{\pot, \text{Local}}}^{(2)}$
as given in \eref{e,c1_pot_local} and \eref{e,c2_pot_local}, respectively,
which can be computationally demanding
because one SVD should be performed for each pair $(\vs, \vp) \in \Lambda \times \Gamma$.
In the following we provide a heuristic fast strategy for this calculation.

Instead of considering every pair $(\vs, \vp) \in \Lambda \times \Gamma$
for calculating
$c_{h_{\pot, \text{Local}}}^{(1)}$ and
$c_{h_{\pot, \text{Local}}}^{(2)}$,
we have empirically observed that it suffices
to use only the pairs associated
with a single shift index,
denoted
$\bar{\vs} \in \Lambda$,
that we typically set to $\vs = \bm{0}$
for simplicity.
For our fast variant,
we approximate the majorizer
$g_{h_{\pot, \text{Local}}}(\alf~; \alfs, \X, \bDel, \gpot)$
by the quadratic function 
\begin{align}
\bar{g}_{h_{\pot, \text{Local}}}(\alf~; \alfs, \X, \bDel, \gpot)
&\defequ
\bar{c}_{h_{\pot, \text{Local}}}^{(0)}(\alfs, \X, \bDel)
+ \bar{c}_{h_{\pot, \text{Local}}}^{(1)}(\alfs, \X, \bDel) \, (\alf-\alfs)
\label{e,maj_hlocal_heu} \\
& \quad
+ \frac{1}{2}\bar{c}_{h_{\pot, \text{Local}}}^{(2)}(\alfs, \X, \bDel, \gpot)
\, (\alf - \alfs)^2,
\nonumber
\end{align}
with coefficients given by:
\begin{align*}
\bar{c}_{h_{\pot, \text{Local}}}^{(0)}(\alfs, \X, \bDel) &=
|\Lambda |\sum_{\vp \in \Gamma}
c_{\hpot}^{(0)}(\alfs, \X^{(\vp, \bar{\vs})}, \bDel^{(\vp, \bar{\vs})}),
\\
\bar{c}_{h_{\pot, \text{Local}}}^{(1)}(\alfs, \X, \bDel) &=
|\Lambda |\sum_{\vp \in \Gamma}
c_{\hpot}^{(1)}(\alfs, \X^{(\vp, \bar{\vs})}, \bDel^{(\vp, \bar{\vs})}),
\\
\bar{c}_{h_{\pot, \text{Local}}}^{(2)}(\alfs, \X, \bDel, \gpot) &=
|\Lambda |\sum_{\vp \in \Gamma}
c_{\hpot}^{(2)}(\alfs, \X^{(\vp, \bar{\vs})}, \bDel^{(\vp, \bar{\vs})}, \gpot).
\end{align*}
This subsequently provides an approximation for the majorizer
$g_{h}(\alf~; \alfs, \X, \bDel, \gpot)$ given by the quadratic function
\begin{align}
\bar{g}_{h}(\alf~; \alfs, \X, \bDel, \gpot) & \defequ g_{h_f}(\alf~; \alfs, \X, \bDel)
+ \reg \bar{g}_{h_{\pot, \text{Local}}}(\alf~; \alfs, \X, \bDel, \gpot)
\nonumber \\
& = \bar{c}_{h}^{(0)}(\alfs, \X, \bDel) + \bar{c}_{h}^{(1)}(\alfs, \X, \bDel) \, (\alf-\alfs)
\nonumber \\
&\quad + \frac{1}{2}\bar{c}_{h}^{(2)}(\alfs, \X, \bDel, \gpot) \, (\alf - \alfs)^2,
\nonumber
\end{align}
with coefficients given by:
\begin{align*}
\bar{c}_{h}^{(0)}(\alfs, \X, \bDel) &= c_{f}^{(0)}(\alfs, \X, \bDel) + \reg 
\bar{c}_{h_{\pot, \text{Local}}}^{(0)}(\alfs, \X, \bDel),
\\
\bar{c}_{h}^{(1)}(\alfs, \X, \bDel) &= c_{f}^{(1)}(\alfs, \X, \bDel) + \reg
\bar{c}_{h_{\pot, \text{Local}}}^{(1)}(\alfs, \X, \bDel),
\\
\bar{c}_{h}^{(2)}(\alfs, \X, \bDel, \gpot) &= c_{f}^{(2)}(\alfs, \X, \bDel) + \reg
\bar{c}_{h_{\pot, \text{Local}}}^{(2)}(\alfs, \X, \bDel, \gpot).
\end{align*}
Then, instead of using \eref{e,ss_iter}, we use the iterative step 
\begin{align}
\alf_{k+1} & \defequ \argmin{\alf \in \reals} \bar{g}_{h}(\alf ~; \alf_k, \X, \bDel, \gpot)
= \alf_{k} - 
\frac{\bar{c}_{h}^{(1)}(\alf_k, \X, \bDel)}{\bar{c}_{h}^{(2)}(\alf_k, \X, \bDel)}.
\label{e,ss_iter_heu}
\end{align}
This heuristic approach is substantially faster than the one in \eref{e,ss_iter}
as only one SVD is performed instead of $|\Lambda||\Gamma|$;
however, a step size produced with this method is not guaranteed
to decrease $h(\alf~; \X, \bDel)$ monotonically,
as $\bar{g}_{h}(\alf~; \alfs, \X, \bDel, \gpot)$
is not necessarily a majorizer for $h(\alf~; \X, \bDel)$.
We have empirically seen that
$\bar{g}_{h}(\alf~; \alfs, \X, \bDel, \gpot)$
resembles $g_{h}(\alf~; \alfs, \X, \bDel, \gpot)$ quite well,
and does not considerably affect performance when estimating \X.
Intuitively, the good performance of this method
could be attributed to some isotropy of the local low-rank regularizer,
in the sense that every shift $\vs\in \Lambda$
contributes similarly to the cost function.

The gradient formula in \eref{e,kost_llr_grad}
and the step-size iterative selection strategy in \eref{e,ss_iter}
allow the minimization of $\kost_{\text{Local}}$
using standard iterative gradient-based optimization algorithms.
In the following section we empirically illustrate this procedure 
in the context of dynamic MRI reconstruction 
using the NCG algorithm; corresponding pseudocode is provided in Algorithm \ref{alg,ncg}.
\begin{algorithm} 
\caption{Nonlinear conjugate gradient algorithm for solving \eref{e,kost_local_h}}
\begin{algorithmic}[1] 
\Require $\X_0 \in \CMN$ (initial estimation), $\alf_0$ (initial step size),
$n_{\alf} \in \mathbb{N}$
(number of iterations for the MM step-size selection strategy),
$M_{\text{iter}}\in \mathbb{N}$ (number of iterations),
$\eta\in \reals$ (convergence threshold)
\State $\G_0 = \nabla \kost_{\text{Local}}(\X_0)$
(initial gradient calculation using \eref{e,kost_llr_grad} and \eref{e,R_grad})
\State $\bDel_0 = -\G_0$ (initial search direction)
\For {$k = 0, 1, \ldots, M_{\text{iter}}-1$}
\State $\alf_k \gets \alf_0$
\For{$l = 1, \ldots, n_{\alf}$} (MM step-size selection)
\State $\alpha_k \gets \alf_k
- \frac{c_{h}^{(1)}(\alf_k, \X_k, \bDel_k)}
{c_{h}^{(2)}(\alf_k, \X_k, \bDel_k, \gpot)}$
(using \eref{e,h_c1} and \eref{e,h_c2})
\EndFor
\State $\X_{k+1} \gets \X_k + \alf_k\bDel_k$ (estimation update)
\State $\G_{k+1} \gets \nabla \kost_{\text{Local}}(\X_{k+1})$
(gradient calculation using \eref{e,kost_llr_grad} and \eref{e,R_grad})
\State $\beta_k \gets \mnormfrob{\G_{k+1}}^2 / \mnormfrob{\G_k}^2$
(conjugate gradient parameter update using Fletcher-Reeves rule)
\State $\bDel_{k+1} \gets -\G_{k+1} + \beta_k \bDel_k$ (search direction update)
\If {$\mnormfrob{\G_{k+1}} < \eta$} 
\State \textbf{break} 
\EndIf 
\EndFor 
\end{algorithmic}
\label{alg,ncg}
\end{algorithm}

\section{Application in dynamic MRI reconstruction using local low-rank constraints}
\label{sec,mri}

In dynamic MRI applications it is typical to acquire a limited amount of samples
to accelerate the acquisition, as data collection is inherently slow in MRI.
Overall, the main goal is to reconstruct a series of images
from a partial set of samples that have also been corrupted by noise.
In the following we assume that these images are 2D,
although our analysis can be easily extended to higher dimensions.
Mathematically, the data-collection process is usually modeled as:
\be
\Y = \jcal{A}(\X) + \bE \in \CSN,
\ee{e,dmri_model}
where
$\X = \begin{bmatrix}
\x_1 & \x_2 & \cdots & \x_N
\end{bmatrix}\in\CMN$
is a matrix whose columns $\{\x_n\}_{n=1}^N$
are the (vectorized) images to be recovered;
$\Y = \begin{bmatrix} \y_1 & \y_2 & \cdots & \y_N \end{bmatrix}\in\CSN$
is a matrix whose columns $\{\y_n\}_{n=1}^N$ are the observed data samples in the Fourier domain
(a.k.a. k-space) for each image;
\bE is a matrix that represents the noise in the observations%
\footnote{The entries in \bE are usually modeled as
independent identically distributed variables
following a white Gaussian complex distribution \cite{liang1999book}.};
and $\jcal{A}:\CMN \arrow \CSN$ is the linear forward operator
that represents the MRI acquisition process%
\footnote{
See \cite{christodoulou2020accelerated} for a detailed explanation of \cA.
In general terms, $\jcal{A}$ involves multiplication with coil-sensitivities,
Fourier transformation, and an undersampling process.}.
The acquisition of a limited amount of samples
is modeled by including an undersampling matrix in $\jcal{A}$.
This makes recovering \X from \Y an ill-posed inverse problem.
One option to recover \X is by minimizing a data-consistency cost function of the form: 
\be
f(\X) \defequ \frac{1}{2}\mnormfrob{\jcal{A}(\X) - \Y}^2,
\ee{e,dc_f}
which can be done using well-known least-squares solvers;
however, additional regularizers are needed in cases when data are heavily undersampled.
In the following we study how to estimate \X
when local low-rank regularizers are added
by using the proposed smooth optimization framework from Section \ref{sec,opt}.
We start by describing how the local low-rank model introduced in \eref{e,kost_local}
and \eref{e,kost_local_h} is applied in the context of dynamic MRI reconstruction. 

\begin{figure}[t] 
\centering 
\includegraphics[width=0.75\textwidth]{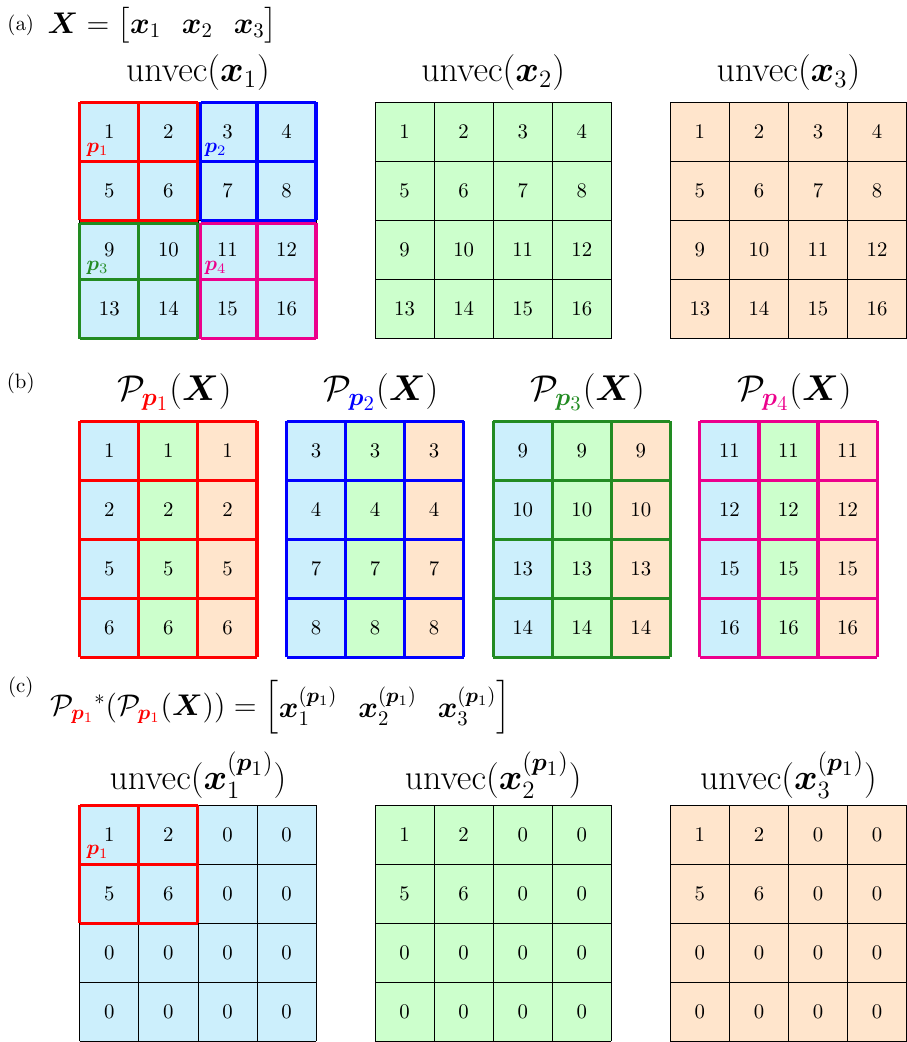} 
\caption{Illustration of the operators $\jcal{P}_{\vp}$ and $\jcal{P}^*_{\vp}$
for a series of $N=3$ 2D-images with $M=16$ voxels each.
(a) shows each image (\ie, the columns of \X) as a 2D grid
after unvectorizing the corresponding column in \X;
the value of each voxel is indicated in black at the center of each square of the grid.
Each image has been filled out with a different color
(light blue, light green, and light orange)
to identify how the voxels are taken from each image
once the Casorati matrices are constructed.
The set of locations
$\Gamma = \{{\color{red}{\bm p_1}}, {\color{blue}{\bm p_2}},
{\color{ForestGreen}{\bm p_3}}, {\color{magenta}{\bm p_4}} \}$
has been indicated in the first image using a color code;
the same locations are used in the other two images.
(b) shows the Casorati matrices created after extracting the patches corresponding
to the locations in $\Gamma$.
A grid with the same color of the location index has been used for each Casorati matrix.
This should help the reader to identify which voxels were taken from each image
to construct the Casorati matrices, 
as the same grid color has been used in (a)
to indicate the neighborhood of voxels that forms the patch corresponding to each location.
In this example the dimensions of each patch were $n_x\times n_y = 2 \times 2$
(\ie, $P = 4$).
(c) shows an example of how $\jcal{P}^*_{\vp}$ performs its matrix construction.
The columns of $\jcal{P}_{{\color{red}{\bm p_1}}}^*(\jcal{P}_{{\color{red}{\bm p_1}}}(\X))$,
denoted by $\{\x_n^{(\bm p_1)}\}_{n=1}^3$, are shown after unvectorization. }
\label{fig:patch}
\end{figure}

\begin{figure}[t] 
\centering 
\includegraphics[width=0.75\textwidth]{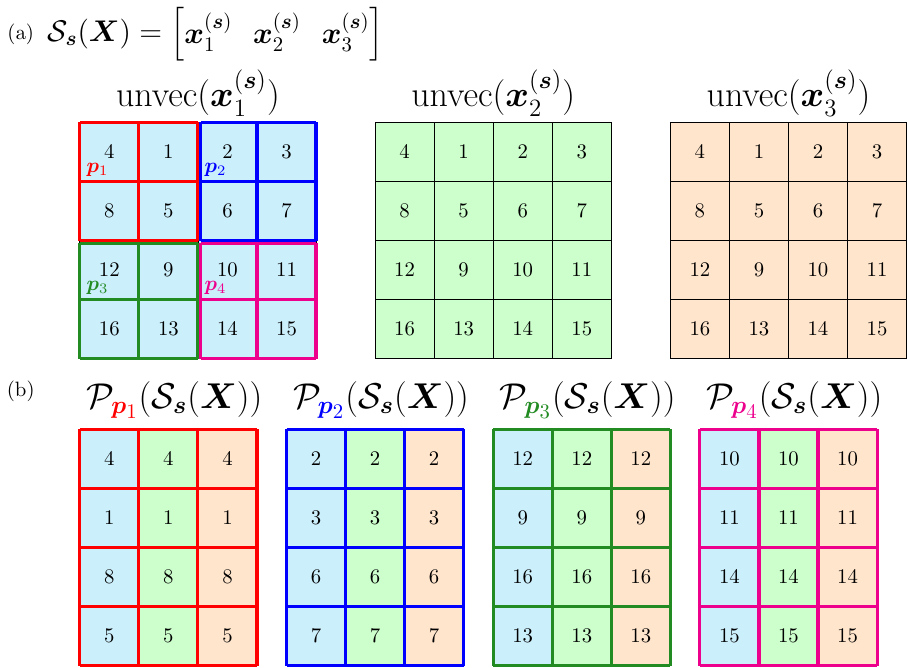} 
\caption{Illustration of the shifting operator
using the same images and color code used in \fref{fig:patch}.
(a) shows how each column of \X is modified
after applying the operator $\jcal{S}_{\vs}$ when $\vs = [1, 0]^T$.
The modified columns are denoted by $\{\x_n^{(\vs)}\}_{n=1}^3$,
and correspond to the vectorized images after circularly shifting the images
according to \vs.
The first and second entries of \vs indicate
how many positions each voxel should be shifted horizontally and vertically, respectively;
in this example $\vs$ indicates one position to the right and no vertical shifts.
The total number of shift vectors depends on the number of voxels in each patch.
In this case
$\Lambda = \{[0, 0]^T, [0, 1]^T, [1, 0]^T, [1, 1]^T\}$
according to \eref{e,shift_set}.
(b) shows the Casorati matrices created
after extracting the patches corresponding to the locations in $\Gamma$
as done in \fref{fig:patch}.b,
after circularly shifting the images according to \vs.}
\label{fig:shift}
\end{figure}

The underlying assumption for the use of local low-rank models
is that patches extracted from the same spatial locations across images should be similar.
More specifically, if $\vp \in \ints^2$ corresponds to the spatial location
of one specific voxel,
it is then possible to extract from each image the values in a neighborhood around $\vp$
corresponding to a patch of dimensions $n_x \times n_y$.
In our examples we used rectangular neighborhoods
and we denote by $P = n_xn_y$ the number of voxels in each of them.
Another possibility is to use ellipsoidal neighborhoods \cite{lobos2022shape}.
Then, low-rank characteristics are expected
for the matrix constructed by concatenating the vectorized patches
corresponding to $\vp$ from the different images.
This matrix has a Casorati structure
and we denote it by $\jcal{P}_{\vp}(\X)$,
where $\jcal{P}_{\vp}:\CMN \arrow \complex^{P\times N}$
is the linear operator that performs the Casorati matrix construction
for the patch corresponding to $\vp$.
The adjoint operator $\jcal{P}^*_{\vp}:\complex^{P\times N}\arrow \CMN$
constructs a matrix that is zero everywhere,
except for the entries corresponding to the patch $\vp$
which are filled according to the entries of the input.
Figure \ref{fig:patch} illustrates how $\jcal{P}_{\vp}$ and $\jcal{P}^*_{\vp}$
perform their respective matrix construction for a series of three images.
Local low-rank models aim to leverage on the low-rank characteristics
of the Casorati matrices constructed from multiple patches at different spatial locations.
As in \eref{e,kost_local} and \eref{e,kost_local_h},
we denote the set of spatial locations by $\Gamma$,
and in our experiments we design it such that
the patches corresponding to different spatial locations
jointly cover the whole voxel grid while not overlapping with each other.
However, one important consideration for local low-rank models to properly work,
is that the patches should overlap
to avoid blocky artifacts in the reconstructed images \cite{saucedo2017improved}.
One way to impose this
is by considering the patch extraction process for circularly shifted versions of the image.
If $\vs \in \ints^2$ denotes a shift index,
then it is expected that the matrix $\jcal{P}_{\vp}(\jcal{S}_{\vs}(\X))$
would possess low-rank characteristics,
where $\jcal{S}_{\vs}:\CMN \arrow \complex^{M\times N}$
is the linear operator that performs the circular shift by \vs.
The adjoint operator $\jcal{S}^*_{\vs}:\CMN \arrow \complex^{M\times N}$
performs a circular shift by $-\vs$.
The number of shifts that we considered depended on the number of voxels in each patch.
If each patch is considered as a 2D grid,
then each location in the grid provides a shift index.
Let $\Lambda$ denote the set of shifts
that also includes
$\vs = [0, 0]^T$ (\ie, no shift),
so that $|\Lambda| = n_x n_y$ if $n_x$ and $n_y$ are both even,
which is the case considered in our experiments.
More specifically,
\be
\Lambda \defequ \left\lbrace
\vs = [s_x, s_y]^T ~ ;
~ s_x \in \left[-\frac{n_x}{2} + 1, \frac{n_x}{2}\right],
~s_y \in \left[-\frac{n_y}{2}+1, \frac{n_y}{2}\right]
\right\rbrace.
\ee{e,shift_set}
Figure \ref{fig:shift} illustrates how the shifting operation
is combined with patch extraction
using the same toy example as in \fref{fig:patch}.
Under this setting, we propose to estimate \X by solving \eref{e,kost_local_h}
using standard iterative gradient-based optimization algorithms
by using the results in \eref{e,kost_llr_grad} and \eref{e,ss_iter}.
As indicated in Section \ref{sec,opt},
we first must calculate $\nabla f(\X)$ and a quadratic majorizer for $h_f(\alf~; \X, \bDel)$.
Given the quadratic structure of $f$ in \eref{e,dc_f} it follows that: 
\be
\nabla f(\X) = \jcal{A}^*(\jcal{A}(\X) - \Y),
\ee{e,grad_f}
and that the coefficients of the quadratic majorizer in \eref{e,hf_maj} are given by:
\begin{align*}
c_{f}^{(0)}(\alfs, \X, \bDel) &=
\frac{1}{2}\mnormfrob{\jcal{A}(\X + \alfs\bDel) - \Y}^2,
\nonumber \\
c_{f}^{(1)}(\alfs, \X, \bDel) &=
\rinprod{ \jcal{A}(\X + \alfs\bDel) - \Y}{\jcal{A}(\bDel)},
\nonumber \\
c_{f}^{(2)}(\bDel) &= \mnormfrob{\jcal{A}(\bDel)}^2.
\nonumber 
\end{align*}

Using the gradient calculation results
in \eref{e,R_grad}, \eref{e,kost_llr_grad}, and \eref{e,grad_f},
in addition to the MM step-size selection strategy in \eref{e,ss_iter}
or its fast (heuristic) version in \eref{e,ss_iter_heu},
we use a method like NCG to solve \eref{e,kost_local_h}.
In the following we show experiments using this optimization approach and dynamic MRI data.

\subsection{Experiments}

In this section we test the proposed optimization framework
by reconstructing realistic retrospectively undersampled dynamic MRI data.
We studied imposing local low-rank models
using the Huber-based low-rank regularizer \rpot and its tail version \rpotk.
The standard iterative gradient-based algorithm used in our experiments was NCG,
which was compared with ad-hoc heuristic versions of two proximal gradients methods:
POGM and FISTA.
For these two proximal gradient methods the cost function was the one in \eref{e,kost_local}.
Since the local low-rank regularizer is not prox friendly,
for both POGM and FISTA
we \emph{approximated}
the proximal mapping for the regularizer
using the average of the proximal mappings of each term inside the sum
(a.k.a. a proximal average \cite{bauschke:08:tpa}).

\subsubsection{Data description and forward operator}
The data used in our experiments corresponded to a 2D cardiac perfusion dataset
of size $128 \times 128$ (\ie, $M = 16384$)
with $N = 40$ time frames \cite{otazo:15:lrp}.
In this case we have data from $12$ coils;
therefore, the forward operator $\jcal{A}$ includes multiplication with coil-sensitivities,
Fourier transformation, and undersampling in the k-space corresponding to each time frame. 
The multicoil k-space data corresponding to each time frame
was undersampled to obtain a $\times7$-accelerated dataset.
The undersampling scheme guaranteed that each k-space sample was present at least once
when considering all time frames.

\subsubsection{Optimization framework parameters}

To use the proposed optimization framework
we first needed a Huber potential function \pot
such that \rpot and \rpotk were able to impose low-rank models.
We focused our analysis on the (convex) hyperbola function defined in \eref{e,pot_hyperbola}
for $\del = 10^{-3}$,
as \rpot and \rpotk in this case
can be seen as smooth versions of the nuclear norm (\cf \fref{fig:pot}.a)
and the regularizer corresponding to the sum of the tail singular values
\cite{oh2016partial}, respectively;
however, we also provide some results using the (non-convex) Cauchy potential function
defined in \eref{e,pot_cauchy}.
In all of our experiments we set $K=1$ for \rpotk.
This value was chosen assuming that,
for this type of data, 
the spatial region covered by each patch
uniformly changes its intensity over time. 
Therefore, each column in the corresponding Casorati matrix of each patch
is considered as a scaled version of any other column,
which produces a rank equal to one. 
To use the MM step-size selection strategy in \eref{e,ss_iter}
we studied $\gpot = \gpotR$ and $\gpot = \gpotL$ when using \rpot,
as both provide valid quadratic majorizers according to Corollary \ref{cr,maj}.
For \rpotk we used the analogous operators
$\gpotw = \gpotwkR$ and $\gpotw = \gpotwkL$ defined in Appendix \ref{s,ap_weight}.
We set the number of iterations for the MM step-size selection strategy
to $n_{\alf} = 1$ and the initial step size as $\alf_0 = 0$.
In cases where the (heuristic) fast step-size selection strategy was used
we set $\bar{\vs} = [0,0]^T$.
The operator $\jcal{P}_{\vp}$ was designed
to extract $8 \times 8$ patches (\ie, $n_x = n_y = 8$),
and we selected $\Gamma$ such that the $128\times 128$ grid of voxels
could be completely covered using non-overlapping patches
before applying the shifting operator.
Thus, $|\Gamma| = (128/n_x)(128/n_y) = 256$ in this example,
and $|\Lambda| = n_xn_y = 64$.
Each method was initialized using a data-sharing reconstruction
\cite{jones:93:kss}
where each time frame was reconstructed by filling missing k-space samples
using available k-space samples in neighboring time frames.
Each method was run for 500 iterations and implemented in-house using MATLAB R2023a.
All computations were performed on a MacBook Pro laptop computer
with an Apple M2 Pro chip, 12 cores, and 32GB RAM.
The authors have made an open-source implementation available at
\url{https://github.com/ralobos/smooth_LLR}.

\subsubsection{Performance metrics}

To study the convergence rate of each method we calculated
the normalized distance to the limit
\be
\frac{\mnormsfrob{\X_k - \hat{\X}_{\infty}}}{\mnormsfrob{\hat{\X}_{\infty}}}
\ee{e,conv_it}
at each iteration, where $\X_k$ was the estimate at the $k$th iteration
and $\hat{\X}_{\infty}$ was the estimate after 500 iterations.
To measure reconstruction quality
we calculated the normalized-root-mean-square-error (NRMSE) at each iteration defined as
\be
\frac{\mnormfrob{\X_k- \X_{\text{GS}}}}{\mnormfrob{\X_{\text{GS}}}},
\ee{e,nrmse_it}
where $\X_{\text{GS}}$ corresponded to the gold-standard reconstruction
obtained after coil-combination of the fully sampled data using a SENSE-based approach
\cite{pruessmann1999sense}.

\begin{figure}[t] 
\centering 
\includegraphics[width=1\textwidth]{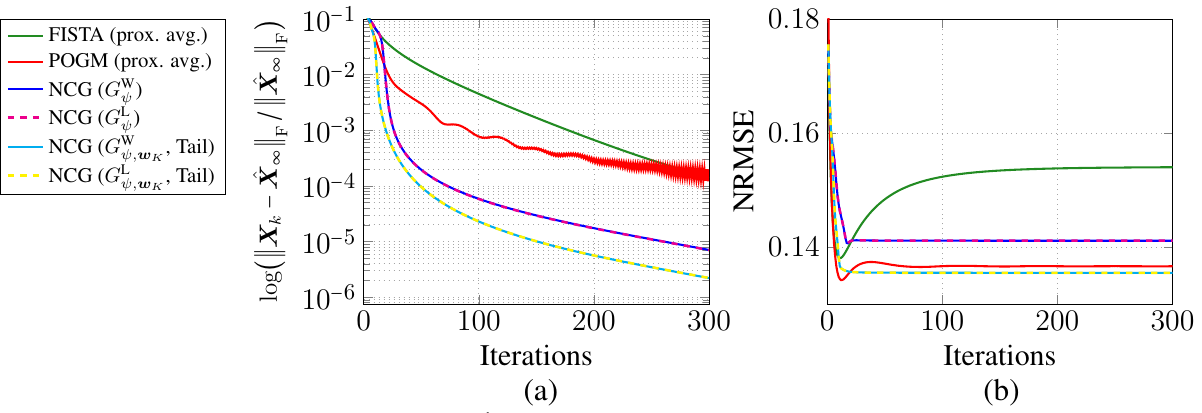}
\caption{Convergence and NRMSE metrics for each method
as functions of the number of iterations (only 300 of the 500 iterations are shown).
In both plots NCG using \rpot corresponds to NCG$(\gpotR)$ and NCG$(\gpotL)$,
and NCG using \rpotk corresponds to NCG$(\gpotwkR, \text{Tail})$
and NCG$(\gpotwkL, \text{Tail})$.
The operator in parenthesis is the one used
for the MM step-size selection strategy in \eref{e,ss_iter}.
For POGM and FISTA we have indicated in parenthesis that proximal average has been used
as an approximation of the proximal mapping.}
\label{fig:conv_nrmse}
\end{figure}

\subsection{Results}

Figure \ref{fig:conv_nrmse} shows the measures
in \eref{e,conv_it} and \eref{e,nrmse_it} for each method.
In \fref{fig:conv_nrmse}.a NCG converged faster than the other two methods,
using either \rpot or \rpotk, and POGM converged faster than FISTA.
The faster convergence of POGM compared to FISTA
was also observed for global low-rank models \cite{lin:19:edp}.
NCG using \rpot yielded similar NRMSE values as POGM,
while taking fewer iterations for NCG to converge.
NCG using \rpotk obtained a better NRMSE than POGM at the point of convergence,
and converged the fastest.
The convergence of NCG is theoretically ensured \cite{hager2006survey}
whereas the convergence of POGM and FISTA and their point of convergence
are unclear in this context
because of the heuristic approximation of the proximal mapping in both cases.
NCG using the tail regularizer \rpotk converged faster and obtained better image quality
than NCG using the non-weighted regularizer \rpot
as expected, because the tail regularizer
exploits a priori estimate of the true rank
and penalizes only tail singular values.

\begin{figure}[t] 
\centering 
\includegraphics[width=1\textwidth]{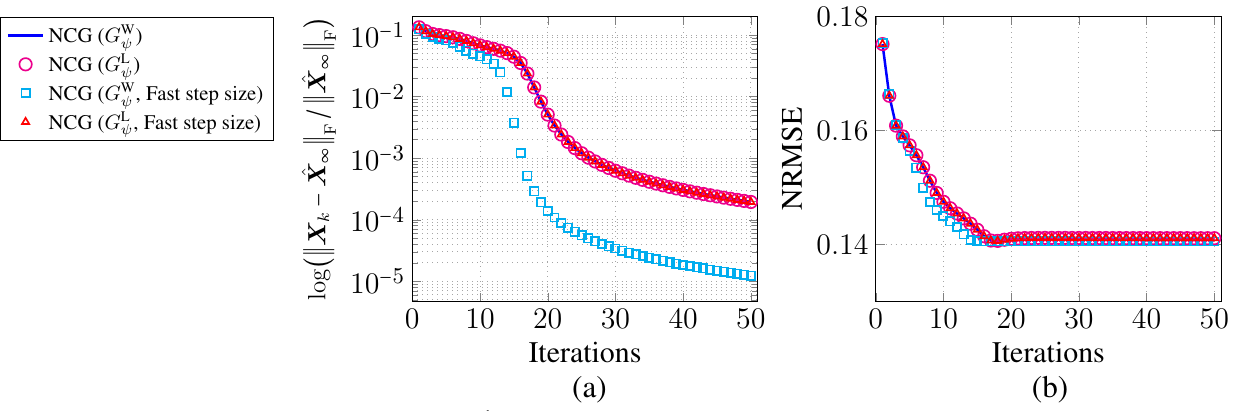}
\caption{Convergence and NRMSE metrics for NCG using \rpot
with different MM step-size selection strategies
as functions of the number of iterations
(only 50 of the 500 iterations are shown).
We indicate using the heuristic step-size selection strategy
by writing ``Fast step size''. }
\label{fig:conv_nrmse_ns}
\end{figure}

\begin{figure}[t] 
\centering 
\includegraphics[width=1\textwidth]{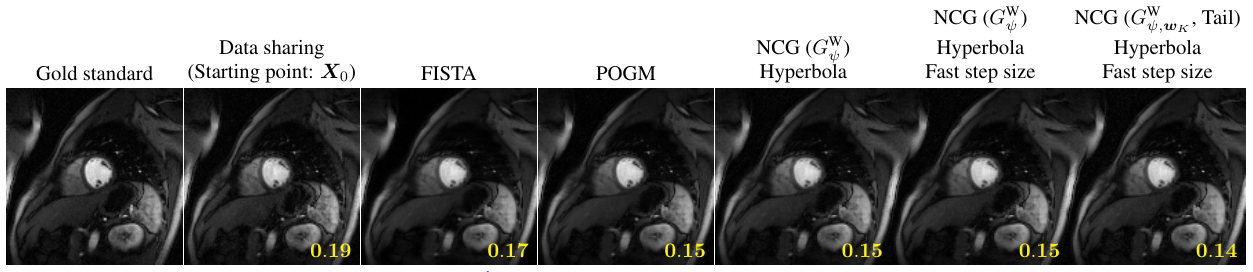}
\caption{Reconstruction results for different methods using only 25 iterations.
Only one image corresponding to one specific time frame is shown for each method,
and the NRMSE for that particular image is indicated in yellow in the bottom right corner.
For NCG using \rpot results are shown using both step-size selection strategies
for a visual comparison between them.
For NCG using \rpotk only the reconstruction
using the heuristic (fast) step-size selection strategy is shown,
as no substantial differences were observed
compared to using the non-heuristic version.
We indicated use of the heuristic step-size selection strategy
by writing ``Fast step size''.
Similar results were obtained for NCG when using \gpotL instead of \gpotR,
and \gpotwkL instead of \gpotwkR (not shown). }
\label{fig:recons}
\end{figure}

\begin{figure}[t] 
\centering 
\includegraphics[width=1\textwidth]{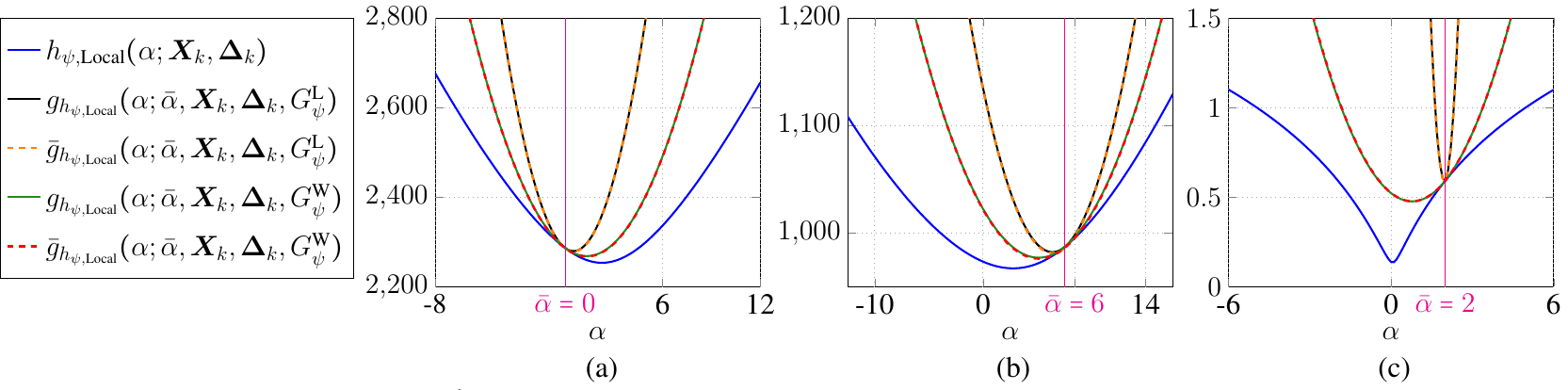}
\caption{Illustration of the proposed quadratic majorizers
for $h_{\pot, \text{Local}}$ given in \tref{th,h_local_maj} when using
(a) \rpot with \pot as the hyperbola function,
(b) \rpotk with \pot as the hyperbola function,
and (c) \rpot with \pot as the Cauchy function.
We also show the approximation of each quadratic majorizer (dashed lines)
as defined in \eref{e,maj_hlocal_heu}.
Different values of \alfs are used in each case
to illustrate different examples of our theoretical results.
When using the tail regularizer the function $h_{\pot, \text{Local}}$
is defined in terms of \rpotk,
and the quadratic majorizers and their approximations are calculated
considering the operators \gpotwkR and \gpotwkL
(\cf Appendix \ref{s,ap_weight}). }
\label{fig:maj_ls}
\end{figure}

Although NCG converged faster than POGM and FISTA,
it was the most computationally expensive,
as many SVD calculations were needed for the MM step-size selection strategy
in \eref{e,ss_iter}.
For this reason we also explored the heuristic step-size selection strategy
proposed in \eref{e,ss_iter_heu}.
Figure \ref{fig:conv_nrmse_ns} shows NCG results
using \rpot and both step-size selection strategies.
Neither convergence nor NRMSE were considerably affected when
adopting the fast (heuristic) step-size selection strategy
(similar results were obtained for NCG using \rpotk).
The heuristic step-size selection did not visibly
affect the quality of the reconstructed images either.
Figure \ref{fig:recons} shows examples of the reconstructed images
obtained with each method using only 25 iterations in each case.
Using the heuristic (fast) step-size selection strategy did not cause visible changes.
However, the fast step-size strategy
significantly reduced NCG compute time
as shown in Table 1
that reports the median times
in all cases after repeating each experiment five times.
Best results are marked in bold. 
Only the results for \gpotR and \gpotwkR are shown,
as no substantial differences were observed when using \gpotL and \gpotwkL instead.
To further study the similarities between the two step-size selection strategies
in terms of convergence and image quality,
we calculated the quadratic functions
$g_{h_{\pot, \text{Local}}}$ and $\bar{g}_{h_{\pot, \text{Local}}}$
for one particular NCG iteration
using \eref{e,h_maj} and \eref{e,maj_hlocal_heu}, respectively,
as they represent the main difference between the two strategies.
Denoting the estimation and search direction for this iteration
as $\X_k$ and $\bDel_k$, respectively,
\fref{fig:maj_ls}.a shows:
$h_{\pot, \text{Local}}(\alf; \X_k, \bDel_k)$
calculated for different values of \alf;
$g_{h_{\pot, \text{Local}}}(\alf~; \alfs, \X_k, \bDel_k, \gpot)$
and $\bar{g}_{h_{\pot, \text{Local}}}(\alf~; \alfs, \X_k, \bDel_k, \gpot)$
when $\alfs =0$ and $\gpot = \gpotR$ or $\gpot = \gpotL$.
As indicated by our theoretical results in \tref{th,h_local_maj},
$g_{h_{\pot, \text{Local}}}(\alf~; \alfs, \X_k, \bDel_k, \gpot)$
majorizes $h_{\text{Local}}(\alf; \X_k, \bDel_k)$ at every \alf
and they are equal at $\alfs$,
both for $\gpot = \gpotR$ or $\gpot = \gpotL$.
Notably,
$\bar{g}_{h_{\pot, \text{Local}}}(\alf~; \alfs, \X_k, \bDel_k, \gpot)$
resembled $g_{h_{\pot, \text{Local}}}(\alf~; \alfs, \X_k, \bDel_k, \gpot)$
quite well for each choice of $\gpot$,
providing further explanation
for why choosing the heuristic step-size selection strategy
did not affect convergence nor estimation quality.
Analogous results are observed in \fref{fig:maj_ls}.b when using the tail regularizer,
although in this case the quadratic functions were calculated based on \gpotwkR and \gpotwkL.
For completeness,
\fref{fig:maj_ls}.c shows an analogous analysis
using the Cauchy function as the potential function for \rpot.
In this case
$h_{\pot, \text{Local}}(\alf; \X_k, \bDel_k)$
has a challenging non-convex behavior;
however, it can still be minimized using the proposed strategy based on quadratic majorizers.
Figure \ref{fig:maj_ls}.a shows that $\gpotR$ provided
better majorizing characteristics than $\gpotL$,
as the global minimum of
$h_{\pot, \text{Local}}(\alf; \X_k, \bDel_k)$ 
was closer to the global minimum of
$g_{h_{\pot, \text{Local}}}(\alf~; \alfs, \X_k, \bDel_k, \gpotR)$
than the global minimum of
$g_{h_{\pot, \text{Local}}}(\alf~; \alfs, \X_k, \bDel_k, \gpotL)$.
This is consistent with the theoretical results in Corollary \ref{cr,maj}.
However, the results in \fref{fig:conv_nrmse} and \fref{fig:conv_nrmse_ns}
illustrate that the differences between the majorizers associated
with $\gpotR$ and $\gpotL$
did not have a big impact when minimizing the cost function in \eref{e,kost_local_h}.
This can be attributed to the fact that
the data-consistency term given by $f$ had a bigger contribution in our experiments
than the regularizer.
As a consequence, the function $g_f$ dominated over $g_{h_{\pot, \text{Local}}}$
when calculating $g_h$ in \eref{e,g_maj} to majorize the line search function $h$,
which mitigated the effects of using $\gpotL$ instead of $\gpotR$. 

\begin{table}[h] 
\centering 
\caption{Reconstruction time and NRMSE after 25 iterations.} 
\begin{tabular}{|c|c|c|} 
\hline
\rowcolor{gray!30}
\textbf{Method} & \textbf{Time [secs]} & \textbf{NRMSE} \\ 
\hline
POGM & \bf{35.1} & 0.138 \\ 
\hline
\rowcolor{gray!30}
NCG $(\gpotR)$ & 75.9 & 0.141 \\ 
\hline
NCG ($\gpotwkR$, Tail) & 76 & \bf{0.136} \\ 
\hline
\rowcolor{gray!30}
NCG ($\gpotR$ ) - Fast step size & 39 & 0.141 \\ 
\hline
NCG ($\gpotwkR$, Tail) - Fast step size & 38.9 & \bf{0.136} \\ 
%
\hline
\end{tabular} 
\end{table}


\clearpage

\section{Discussion and conclusion} \label{sec,dis}

In this work we have proposed new classes of smooth regularizers
to impose either global or local low-rank models in inverse problems.
We have studied several properties of these regularizers
using singular value function theory, establishing conditions for their convexity,
differentiability, and Lipschitz continuity of the gradient.
We have also provided novel theory that enables the construction of majorizers
for the proposed regularizers,
which corresponds to one of our main theoretical contributions.
Moreover, we have shown that these regularizers enable the use of
standard iterative gradient-based optimization methods
to solve inverse problems using low-rank models.
For this purpose, we have shown how to calculate the regularizer gradient in closed-form,
and also how to efficiently
select the step-size in each iteration
using a MM strategy
based on quadratic majorizers of the regularizer line search function. 

Another point to highlight is the flexibility offered by the proposed theoretical framework,
in the sense that a big family of low-rank regularizers can be analyzed
using the same theoretical results.
Many Huber-like potential functions have been proposed
and each of them would provide a regularizer according to our theory.
Interestingly, some of these Huber-based low-rank regularizers
can be considered smooth versions of commonly used non-smooth low-rank regularizers.
For example, the nuclear norm can be approximated 
by selecting a potential function like the hyperbola function
that approximates the absolute value function,
and non-convex Schatten-$p$ quasi-norms with $0<p<1$ can be approximated
by selecting the potential function to be
\(
\pot(t) = (|t|^2 + \delta^2)^{p/2}
\)
for a small value of $\delta$.
This same principle holds
for the proposed smooth weighted Huber-based low-rank regularizers
(\cf Section \ref{sec,weighted_reg}).
Remarkably, the proposed optimization framework
enables the minimization of each Huber-based low-rank regularizer
using the same iterative gradient-based optimization algorithm. 

As an application,
we showed that the proposed optimization framework
based on smooth low-rank regularizers
can be used to address the dynamic MRI reconstruction problem
when using local low-rank models with overlapping patches.
Particularly, we showed that NCG can be used to minimize the associated cost function.
NCG showed a faster convergence
and similar or better image quality
than other heuristic ad-hoc proximal methods
that use the nuclear norm as a regularizer.
Moreover, methods like NCG provide well-known convergence guarantees \cite{hager2006survey}
that are not available for proximal methods with approximate proximal maps in this context.
In addition,
we have observed (not shown) that quasi-Newton methods like L-BFGS
can also be used instead of NCG,
providing similar reconstruction quality. 

As future work we would like to explore other applications
for the proposed tail Huber-based low-rank regularizers studied in Section \ref{sec,weighted_reg}.
Interestingly, by selecting the Huber potential function as the hyperbola function,
the proposed regularizer can be seen as a smooth version
of the regularizer corresponding to the sum of the tail singular values,
which has been used in robust PCA applications \cite{oh2016partial,cavazos2025alpcah}.
Similarly, by selecting the Huber potential function as the parabola,
this type of regularizer coincides with the regularizer proposed in \cite{haldar:14:lrm},
which has been extensively used in MRI reconstruction applications
using structured low-rank models
\cite{haldar2016p,kim2017loraks,lobos2018navigator,bilgic2018improving}.
The theory and optimization framework proposed in this work
could allow a complementary analysis for these applications. 

From a theoretical point of view a relevant question is
whether ``better'' majorizers can be constructed for the proposed regularizers.
The majorizer $Q(\X; \S, \gpot)$ proposed in \eref{e,maj_reg_gen}
depends on the operator $\gpot$,
and \tref{th,maj_Huber} shows that two choices for this operator provide valid majorizers.
However, we have not studied if these majorizers satisfy any optimality criterion.
Inspired by the criterion used to study the optimality of quadratic majorizers
for Huber potential functions
\cite[p.~186]{huber:81},
we would like to study the existence and construction of an operator $\gpot^*$ such that:
\begin{enumerate}[(i)]
\item $Q(\X; \S, \gpot^*)$ is a valid majorizer for \rpot and,
\item $Q(\X; \S, \gpot^*)\leq Q(\X; \S, \gpot), ~ \forall \X\in \CMN$,
\raggedright
for any operator $\gpot$ that makes $Q(\X; \S, \gpot)$ a valid majorizer for \rpot.
\end{enumerate}
Under this criterion the operator \gpotR
defined in \eref{e,gpot_R_wide} and \eref{e,gpot_R_tall}
would be better than the operator \gpotL
defined in \eref{e,gpot_L_wide};
however, it is an open question if there is a better operator than \gpotR.
Answering this is important, as it would allow us to define better quadratic majorizers
for the regularizer line search function and,
therefore, it would improve our proposed majorize-minimize step-size selection strategy. 
\section*{Acknowledgments}
The authors thank
Douglas Noll for discussions about MRI
and Caroline Crockett
for inspiration for some points.

\begin{appendix}

\section{Proof of \tref{th,cvx_dif_R}} \label{s,ap_proof_prop_dif} 

We first prove (i).
Suppose \pot is convex.
Then for $\x, \y \in\reals^r$ and $\alf \in (0,1)$:
\begin{align*}
f_{\pot}(\alf\x + (1- \alf)\y) &= \sum_{k=1}^r\pot(\alf x_k + (1 - \alf)y_k)
\\
&\leq \sum_{k=1}^r\alf\pot(x_k) + (1-\alf)\pot(y_k)
= \alf f_{\pot}(\x) + (1-\alf)f_{\pot}(\y),
\end{align*}
using convexity of \pot in the inequality.
Thus \fpot is convex
and hence \rpot is convex
by Proposition~\ref{th,cvx_dif_F} part~(i)
using the equivalent formulation for \rpot provided in \eref{e,reg_comp}.

Conversely,
if \rpot is convex
then Proposition \ref{th,cvx_dif_F} part (i)
shows that \fpot is convex.
Because $\pot(t) = \frac{1}{r}\fpot(t\vone_r)$
where $\vone_r\in\reals^r$ is the vector with each entry equal to one,
it follows that for $\alf\in (0,1)$ and $t,s\in \reals$: 
\begin{align*}
\pot(\alf t + (1-\alf)s) & = \frac{1}{r}\fpot(\alf t\vone_r + (1-\alf)s\vone_r)
\\
& \leq \alf\frac{1}{r}\fpot(t\vone_r) + (1-\alf)\frac{1}{r}\fpot(s\vone_r)
= \alf \pot(t) + (1-\alf)\pot(s),
\end{align*}
using convexity of \fpot in the inequality.

To prove (ii) we first calculate $\nabla f_{\pot}(\vsig(\X))$.
By definition of $f_{\pot}$,
for any $\x\in\reals^r$:
\begin{align*}
\nabla f_{\pot}(\x) &= [\dpot(x_1),\ldots,\dpot(x_r)]^T
= \dpotdot{\x},
\end{align*}
as \pot is differentiable.
Now we show that \fpot is locally Lipschitz,
which follows from showing that $\nabla\fpot$ is Lipschitz continuous.
If $\x,\y \in \reals^r$, then 
\begin{align*}
\norm{\nabla \fpot(\x) - \nabla \fpot(\y)}^2 & =
\sum_{k=1}^r |\dpot(x_k) - \dpot(y_k)|^2
\\
& \leq 
\sum_{k=1}^r \wpot(0)^2 |x_k - y_k|^2
= \wpot^2(0) \norm{\x- \y}^2,
\end{align*}
using the fact that \dpot is $\wpot(0)$-Lipschitz continuous
(\cf, \eref{e,dpot_lips}) in the inequality.
Therefore, $\nabla\fpot$ is $\wpot(0)$-Lipschitz continuous
which implies that \fpot is locally Lipschitz.
The result
\eqref{e,R_grad}
follows from Proposition \ref{th,cvx_dif_F} part (ii) and (iii)
using the equivalent formulation for \rpot provided in \eref{e,reg_comp}.

\section{Proof of \tref{th,maj_Huber}}
\label{th,proof_maj_Huber}

It is clear from their definitions that
$Q(\cdot \, ; \S, \gpotR)$ and $Q(\cdot \, ; \S, \gpotL)$
both satisfy the equality \eref{e,maj_a_reg}.
Next, we prove that $Q(\cdot \, ; \S, \gpotR)$ satisfies \eref{e,maj_b_reg},
which is the main part of the proof.
We then prove that $Q(\cdot \, ; \S, \gpotL)$ satisfies \eref{e,maj_b_reg}
by showing that
$Q(\X;\S, \gpotR) \leq Q(\X;\S, \gpotL), ~ \forall \X\in\CMN$. 
Without loss of generality we assume 
that $M<N$ (wide case).

A key step in our proof uses the following inequality
to lower bound the trace of a product of two Hermitian matrices
\cite[Thm.~4.3.53]{horn2012matrix}.

\begin{proposition}
[Lower and upper bounds for the trace of the product of two Hermitian matrices]
Let $\A \in \CMM$, $\B\in\CMM$ be two Hermitian matrices 
with eigenvalues 
$\{\lam_k(\A)\}_{k=1}^M$ and $\{\lam_k(\B)\}_{k=1}^M$, respectively. 
Assuming both sets of eigenvalues are arranged in nonincreasing order, \ie,
$\lam_1(\A)\geq\lam_2(\A)\geq\cdots\geq \lam_M(\A)$ and 
$\lam_1(\B)\geq\lam_2(\B)\geq\cdots\geq \lam_M(\B)$, it follows that:
\be
\sum_{k = 1}^M \lam_{M-k+1}(\A) \, \lam_{k}(\B)
\leq \trace{\A\B}
\leq \sum_{k = 1}^M \lam_k(\A) \, \lam_k(\B).
\ee{e,trace_prod}
\end{proposition}

Equipped with this inequality we now continue our proof.
We start showing the inequality \eref{e,maj_b_reg} when $\S\in\CMN$
is a rectangular diagonal matrix of the form $\S = \diag(\vs)$,
where $\vs \in \reals^M$ has nonnegative entries
arranged in nonincreasing order (\ie, $\vs = \check{\vs}$).
We then generalize the result to generic rectangular matrices
by leveraging the unitary invariance of \rpot. 

An important result used in our proof is given in the following lemma.
\begin{lemma}\label{l,qpot_dif}
Let $t,s, \in \reals$. Then,
\be
\pot(s) - \pot(t) + \frac{1}{2}\wpot(s) \, (t^2-s^2) \geq 0.
\ee{e,maj_geq_1d}
\end{lemma}
\begin{proof}
Because $q_{\pot}$ is a majorizer for \pot:
\begin{align*}
0 \leq
q_{\pot}(t;s) - \pot(t)
&=
\pot(s) - \pot(t) + \dpot(s)(t-s) + \frac{1}{2}\wpot(s) \, (t-s)^2
\\ &=
\pot(s) - \pot(t) + s \wpot(s) (t-s) + \frac{1}{2} \wpot(s) \, (t^2 - 2 t s + s^2)
\\ &=
\pot(s) - \pot(t) + \frac{1}{2} \wpot(s) \, (t^2 - s^2)
.\end{align*}
\end{proof}

Next we extend Lemma \ref{l,qpot_dif}
to prove that
$Q(\X ; \S, \gpotR) - \rpot(\X) \geq 0,~ \forall \X\in\CMN$
when $\S = \diag{\vs}$.
In this case we have that $\S = \U\diag{\vsig(\S)}\V'$
where $\vsig(\S) = \vs$, $\U = \I_M$, $\V = \I_N$,
and \I_K denotes the $K\times K$ identity matrix.
This implies that $\nabla\rpot(\S) = \diag{\dpotdot{\vs}}$
by \tref{th,cvx_dif_R}.
It follows that:
\begin{align}
Q(\X ; \S, \gpotR) &= \rpot(\diag{\vs})
+ \rInprod{\diag{\dpotdot{\vs}}}{\X -\diag{\vs}}
\\&\quad
+ \frac{1}{2} (\vone_M' (\gpotR(\vs) \odot |\X -\diag{\vs}|.^{\wedge}2) \vone_N)
\nonumber\\
& = \sumkm \pot(s_k) + \real{\sumkm \dpot(s_k) \, (X_{kk} - s_k)}
\\&\quad
+ \frac{1}{2} \sumkm \wpot(s_k) |X_{kk} - s_k|^2
+ \frac{1}{2} \sumkm \wpot(s_k) \sum_{\substack{l=1 \\ l\neq k}}^N |X_{kl}|^2. 
\nonumber\\
& = \sumkm \pot(s_k) + s_k \wpot(s_k) \, (\real{X_{kk}} - s_k)
\nonumber\\
& \quad + \frac{1}{2} \wpot(s_k) (|X_{kk}|^2 - 2 s_k \real{X_{kk}} + s_k^2)
+ \frac{1}{2} \sum_{\substack{l=1 \\ l\neq k}}^N \wpot(s_k) |X_{kl}|^2
\nonumber\\
& = \sumkm \pot(s_k)
+ \frac{1}{2} \wpot(s_k) (|X_{kk}|^2 - s_k^2)
+ \frac{1}{2} \sum_{\substack{l=1 \\ l\neq k}}^N \wpot(s_k) |X_{kl}|^2
\nonumber\\
& = \left( \sumkm \pot(s_k) - \frac{1}{2} \wpot(s_k)s_k^2 \right)
+ \frac{1}{2} \sumkm \wpot(s_k) \sum_{l=1}^N |X_{kl}|^2.
\label{e,maj_proof_bb}
\end{align}

The next step is to derive a lower bound
for the last sum of the previous equation in terms of the singular values of \X.
For this we first write this term
as the trace of the product of two Hermitian matrices
and use \eref{e,trace_prod} as follows:
\begin{align*}
\sumkm \wpot(s_k)\sum_{l=1}^N |X_{kl}|^2
& = 
\trace{\left(\diag{\sqrt{\wpotdot\vs}}\right)'\X
\left(\left(\diag{\sqrt{\wpotdot\vs}}\right)'\X\right)'}
\\& = \trace{\underbrace{\diag{\sqrt{\wpotdot\vs}}
\left(\diag{\sqrt{\wpotdot\vs}}\right)'}_{\A\in\CMM}
\underbrace{\X\X'}_{\B\in\CMM}}
\\
& = \trace{\A\B}
\geq\sumkm \lam_{M-k+1}(\A)\lam_k(\B).
\end{align*}
We have that $\lam_{M-k+1}(\A) = \wpot(s_k)$
given that \wpot is monotone nonincreasing,
and that $\lam_k(\B) = \sig_k^2(\X)$.
It follows that: 
\begin{align}
\sumkm \wpot(s_k) \sum_{l=1}^N |X_{kl}|^2 
& \geq
\sumkm \wpot(s_k) \sigma^2_k(\X).
\end{align}
Using this inequality to lower bound the right-hand-side in \eref{e,maj_proof_bb},
Lemma \ref{l,qpot_dif} shows
that: 
\begin{align}
Q(\X ; \S, \gpotR) - \rpot(\X) &\geq \sumkm \pot(s_k) - \pot(\sigma_k(\X))
+ \frac{1}{2}\wpot(s_k)(\sigma^2_k(\X) - s_k^2)
\geq 0.
\label{e,Q-R,diag}
\end{align}

Now we prove the general case.
Let $\Y, \S \in \CMN$
and $\U\in \complex^{M\times M}, \V\in\complex^{N\times N}$
denote two unitary matrices such that $\S = \U\diag{\vsig(\S)}\V'$.
Because \rpot is unitarily invariant, it follows that:
\begin{align}
\rpot(\Y + \S) & = \rpot(\U'(\Y + \S)\V) 
= \rpot(\U'\Y\V + \diag{\vsig(\S)}) 
\nonumber 
\\
&\leq Q(\U'\Y\V + \diag{\vsig(\S)} ; \diag{\vsig(\S)}, \gpotR) 
\quad \text{ by \eref{e,Q-R,diag} }
\nonumber 
\\
&= \rpot(\diag{\vsig(\S)}) + \rinprod{\nabla \rpot(\diag{\vsig(\S)})}{\U'\Y\V}
\nonumber 
\\
&\quad + \frac{1}{2}(\vone_M'(\gpotR(\vsig(\S)) \odot |\U'\Y\V|.^{\wedge}2) \vone_N)
\quad \text{ by \eqref{e,maj_reg_gen} }
\nonumber 
\\
& = \rpot(\S) + \rInprod{\diag{\dpotdot{\vsig(\S)}}}{\U'\Y\V}
\quad \text{ by \eqref{e,R_grad} }
\nonumber 
\\
&\quad + \frac{1}{2}(\vone_M'(\gpotR(\vsig(\S)) \odot |\U'\Y\V|.^{\wedge}2) \vone_N) 
\nonumber 
\\
& = \rpot(\S) + \real{\trace{\diag{\dpotdot{\vsig(\S)}}\V'\Y'\U}}
\nonumber 
\\
& \quad + \frac{1}{2}(\vone_M'(\gpotR(\vsig(\S)) \odot |\U'\Y\V|.^{\wedge}2) \vone_N) 
\nonumber
\\
& = \rpot(\S) + \real{\trace{\U\diag{\dpotdot{\vsig(\S)}}\V'\Y'}}
\nonumber 
\\
& \quad + \frac{1}{2}(\vone_M'(\gpotR(\vsig(\S)) \odot |\U'\Y\V|.^{\wedge}2) \vone_N) 
\nonumber \\
& = \rpot(\S) + \rinprod{\nabla\rpot(\S)}{\Y}
+ \frac{1}{2}(\vone_M'(\gpotR(\vsig(\S)) \odot |\U'\Y\V|.^{\wedge}2) \vone_N),
\label{e,maj_end}
\end{align}
%
using the cyclic property of the trace in the next to last equality.
The result follows from setting $\Y$ to $\X - \S$ in \eref{e,maj_end}.

To show that $Q(\X;\S, \gpotR) \leq Q(\X;\S, \gpotL), ~ \forall \X\in\CMN$,
it is enough to show that
$\vone_M'(\gpotR(\vsig(\S)) \odot |\U'(\X - \S)\V|.^{\wedge}2) \vone_N
\leq \vone_M'(\gpotL(\vsig(\S)) \odot |\U'(\X - \S)\V|.^{\wedge}2) \vone_N$,
which is equivalent to showing that: 
\begin{align*}
\sum_{k=1}^M \wpot(\sigma_k(\S)) \sum_{l=1}^N |[\U'(\X -\S)\V]_{kl}|^2
&\leq \sum_{k=1}^M \left( \sup_{t\in\reals} \wpot(t) \right)
\sum_{l=1}^N |[\U' (\X - \S) \V]_{kl}|^2
\\
& = \sum_{k=1}^M \sum_{l=1}^N \wpot(0) |[\U'(\X -\S)\V]_{kl}|^2,
\end{align*}
where we have used the property that
$\sup_{t\in\reals}\wpot(t) = \wpot(0)$ as \pot is even
and \wpot(t) is monotonic nonincreasing for $t>0$.

The proof for the tall case, \ie, when $M>N$, is analogous
and is omitted for the sake of space.
Finally we prove
\eref{e,maj_reg_gen_R_frob_w},
\eref{e,maj_reg_gen_L_frob_w},
and
\eref{e,maj_reg_gen_R_frob_t}.
To prove \eref{e,maj_reg_gen_R_frob_w}
it is enough to show that:
$$
\mnormfrob{\W_{\pot, \S}'(\X -\S)}^2
= \vone_M'(\gpotR(\vsig(\S)) \odot |\U'(\X - \S)\V|.^{\wedge}2) \vone_N.
$$
Because the Frobenius norm is unitarily invariant we have that:
\begin{align*}
\mnormfrob{\W_{\pot, \S}' (\X -\S)}^2
& =
\mnormfrob{\V' \left(\diag{\sqrt{\wpotdot{\vsig(\vs)}}}\right)' \U' (\X -\S)}^2
\\ & =
\mnormfrob{ \left(\diag{\sqrt{\wpotdot{\vsig(\vs)}}}\right)'  \U' (\X -\S) \V}^2
\\ & =
\sum_{k=1}^M \wpot(\sigma_k(\S)) \, \sum_{l=1}^N |[\U'(\X -\S)\V]_{kl}|^2
\\ & =
\vone_M' (\gpotR(\vsig(\S) \odot |\U'(\X - \S)\V|.^{\wedge}2) \vone_N.
\end{align*}
The proofs for \eref{e,maj_reg_gen_R_frob_t}  and 
\eref{e,maj_reg_gen_L_frob_w} are analogous to the previous proof
and are omitted.

\section{Proof of proposition \ref{th,smooth_R}}
\label{th,proof_smooth_R}
Let $\X, \S \in \CMN$, and $\U, \V$ denote suitable sized unitary matrices
such that $\S = \U\diag{\vsig(\S)}\V'$.
Using \eref{e,maj_reg_gen_L_frob_w} in \tref{th,maj_Huber}:
\be
 \rpot(\X) \leq \rpot(\S) + \rinprod{\nabla \rpot(\S)}{\X -\S}
+ \frac{1}{2} \wpot(0) \mnormfrob{\X -\S}^2, 
\ee{e,maj_lips_proof}
which is equivalent to $\nabla\rpot$ being $\wpot(0)$-Lipschitz
\cite[Thm.~5.8]{beck:17:fom}
for the inner product in \eref{e,inner_prod}.

\section{Proof of \tref{th,hpot}}
\label{th,hpot_proof}
We need to verify that
\begin{eqnarray}
h_{\pot}(\alfs; \X, \bDel) &=& g_{\hpot}(\alfs ; \alfs, \X, \bDel, \gpot),
\label{e,maj_a_hr}\\
h_{\pot}(\alf; \X, \bDel) &\leq& g_{\hpot}(\alf ; \alfs, \X, \bDel, \gpot),
\quad \forall \alf\in \reals.
\label{e,maj_b_hr}
\end{eqnarray}
\eref{e,maj_a_hr} is easy to verify from the definition of
$g_{\hpot}(\cdot ~ ; \alfs, \X, \bDel, \gpot)$.
To prove \eref{e,maj_b_hr} we use the fact that
$Q(\cdot ~; \X + \alfs\bDel, \gpot)$ is a majorizer for \rpot at $\X + \alfs\bDel$.
We have that: 
\begin{align*}
h_{\pot}(\alf;\X, \bDel) & = \rpot(\X + \alf \bDel)
\\ & \leq
Q(\X + \alf \bDel ~; \X + \alfs\bDel, \gpot)
\\ & =
\rpot(\X + \alfs\bDel)
+ \rinprod{\nabla \rpot(\X + \alfs\bDel)}{\X + \alpha\bDel - (\X + \alfs\bDel)}
\\ & \quad
+ \frac{1}{2}\vone_M'(\gpot(\vsig(\X + \alfs\bDel)) \odot
(\U'(\X + \alpha\bDel - (\X + \alfs\bDel))\V).^{\wedge}2) \vone_N
\\ & =
\rpot(\X + \alfs\bDel) + \rinprod{\nabla \rpot(\X + \alfs\bDel)}{\bDel}(\alf - \alfs)
\\ & \quad
+ \frac{1}{2}\vone_M'(\gpot(\vsig(\X + \alfs\bDel)) \odot
(\U'\bDel\V).^{\wedge}2) \vone_N (\alf - \alfs)^2.
\\ & =
g_{\hpot}(\alf ; \alfs, \X, \bDel, \gpot)
, \quad \forall \alf, \alfs \in \reals.
\end{align*}

\section{Proof of Corollary \ref{cr,maj}}
\label{cr,maj_proof}
Without loss of generality we assume that $M<N$ (wide case).
The first part of the corollary follows directly
from \tref{th,maj_Huber} and \tref{th,hpot}.
To show the second part it is enough to show that
$\vone_M'(\gpotR(\vsig(\X + \alfs\bDel)) \odot |\U'\bDel\V|.^{\wedge}2) \vone_N
\leq \vone_M'(\gpotL(\vsig(\X + \alfs\bDel)) \odot |\U'\bDel\V|.^{\wedge}2) \vone_N$.
This is equivalent to showing that:
\begin{align*}
\sum_{k=1}^M \wpot(\sigma_k(\X + \alfs\bDel)) \sum_{l=1}^N |[\U'\bDel\V]_{kl}|^2
&\leq \left(\sup_{t\in\reals}\wpot(t)\right) 
\sum_{k=1}^M \sum_{l=1}^N  |[\U'\bDel\V]_{kl}|^2
\\ & =
\wpot(0) \sum_{k=1}^M \sum_{l=1}^N |[\U'\bDel\V]_{kl}|^2.
\end{align*}

\section{Proof of Proposition \ref{prop,fpotc_cvx}}
\label{ap,proof_fpot_cvx}
We first prove the sufficient condition
by providing an equivalent expression
for \fpotc
in \eref{e,fpotc}
inspired by 
the c-spectral norm
in \cite{lewis1995convex}.
For this purpose
we review \textit{generalized permutation matrices}.

\begin{definition}[Generalized permutation matrix~\cite{lewis1995convex}]
A $r\times r$ matrix \P is a generalized permutation matrix
if it has exactly one nonzero entry in each row and each column,
that entry being equal to $1$ or $-1$.
We denote the set of $r\times r$ generalized permutation matrices by $\Theta_r$.
\end{definition}

Using the previous definition we first show that: 
\be
\fpotc(\x) = \max_{\P \in \Theta_r} \inprod{\w}{\potdot(\P\x)},
\ee{e,f_inner_max}
where here $\inprod{\cdot}{\cdot}$
denotes the Euclidean inner product in $\reals^r$.
Given that $\check{\x}$ is constructed by sorting the entries of $|\x|$,
there exists $\check{\P} \in \Theta_r$ such that $\check{\x} = \check{\P}\x$.
Then, from \eref{e,fpotc}:
\begin{align*}
\fpotc(\x) & =
\sumno_{k = 1}^r w_k \pot([\check{\P}\x]_k)
= \inprod{\w}{\potdot(\check{\P}\x)}
\leq \max_{\P \in \Theta_r} \inprod{\w}{\potdot(\P\x)}.
\end{align*}
We prove the reverse inequality by contradiction. 
If the inequality is not true, 
then there is an \x and
a generalized permutation matrix $\P^* \in\Theta_r$ 
such that 
$\P^*\x \neq \check{\x}$
and
$$
0 \leq
\fpotw(\x) < \inprod{\w}{\potdot(\P^*\x)}
= \sumno_{k=1}^r w_k \pot([\P^*\x]_k).
$$
Because
$\P^*\x \neq \check{\x}$,
there are entries $k_1, k_2\in\{1,\ldots, r\}$
where $k_1<k_2$, 
such that $|[\P^*\x]_{k_1}| < |[\P^*\x]_{k_2}|$. 
Because the weights are nonincreasing,
and $\pot(t)$ is even and nondecreasing for $t>0$,
it follows that 
$$
w_{k_1} \pot(|[\P^*\x]_{k_2}|) \geq w_{k_2} \pot(|[\P^*\x]_{k_1}|).
$$
Let \Pkk denote the generalized permutation matrix 
that swaps the $k_1$th and $k_2$th entries
and multiplies each entry either by 1 or -1 to make both entries nonnegative.
Then we have that  
$$
\inprod{\w}{\potdot(\P^*\x)} 
= \sum_{k=1}^r w_k \pot([\P^*\x]_k) 
\leq \sum_{k=1}^r 
w_k\pot([\Pkk \P^* \x]_k)
= \inprod{\w}{\potdot(\Pkk \P^* \x)}.
$$
At this point either $\Pkk \P^* \x = \check{\x}$ 
or we can find another pair of entries in $\Pkk \P^* \x$ 
that can be swapped and made nonnegative
to majorize $ \inprod{\w}{\potdot(\Pkk \P^* \x)}$.
Continuing this process we finally obtain that 
$\fpotw(\x) < \inprod{\w}{\potdot(\P^*\x)} \leq \fpotw(\x)$,
which is a contradiction. Therefore,
$$
\fpotw(\x) \geq \max_{\P \in \Theta_r} \inprod{\w}{\potdot(\P\x)}.
$$

It is easy to show that each function
$\gam_{\pot}(\x; \w, \P) \defequ \inprod{\w}{\pot.(\P\x)}, ~\P\in \Theta_r $
is convex when \pot is convex.
Then, \fpotc is convex as it can be written as the pointwise supremum
of a collection of convex functions \cite[Thm.~5.5]{rockafellar:70}.

\newcommand{\bmat}[1] {\left[ \begin{matrix} #1 \end{matrix} \right]}

We now prove the necessary conditions.
Noting that
\(
\pot(t) =
\fpotw(\vone_r t) / \sum_{k=1}^r w_k
,\)
convexity of \pot
follows from convexity of \fpotw
because convexity is preserved under affine maps
and nonnegative scaling.

Now we show that the entries of \w must be 
nonincreasing,
using proof by contradiction.

Consider first the case where $r = 2$
with
$0 \leq w_1 < w_2$
and define
$\x = \bmat{u \\ v}$
and
$\y = \bmat{v \\ u}$
for
\(
0 < u < v < \eps \, u 
,\)
where 
$\eps > 1$.
In addition, we pick 
$u$ 
and 
$v$
such that
$\wpot(\alf v + (1-\alf)u) > 0, 
~ \forall \alf \in (0,1)$.
This is possible as \wpot is not always equal to 0 
unless \pot is a constant, 
which is precluded by our ongoing assumptions.
Define
\begin{align*}
g(\alf) & \defequ
\fpotw(\alf \x + (1-\alf) \y)
= \fpotw\of{ \bmat{\alf u + (1-\alf) v \\ \alf v + (1-\alf) u} }
\\&
= \leftbrace{
w_1 \pot(\alf u + (1-\alf) v) + w_2 \pot(\alf v + (1-\alf) u)
, & 0 \leq \alf \leq 1/2
\\
w_1 \pot(\alf v + (1-\alf) u) + w_2 \pot(\alf u + (1-\alf) v)
, & 1/2 \leq \alf \leq 1.
}
\end{align*}
We will show that $g$ is nonconvex,
which means that \fpotw must be nonconvex,
by showing that $g$ increases on $(0,1/2)$
and then decreases on $(1/2,1)$.

If $\alf \in (0,\frac{1}{2})$, then 
$\dot{g}(\alf)$, \ie,
the derivative of $g$ with respect to \alf, satisfies:
\begin{align*}
    \dot{g}(\alf) & =
    w_1 \dpot(\alf u + (1-\alf) v) \, (u - v)
    + w_2 \dpot(\alf v + (1-\alf) u) \, (v - u)
    \\&
    = (v - u) \, \big(w_2 \, \dpot(\alf v + (1-\alf) u)
    -  w_1 \, \dpot(\alf u + (1-\alf) v) \big)
    \\&
    = (v - u) \, \big(w_2 \, (\alf v + (1-\alf) u) \, \wpot(\alf v + (1-\alf) u)
    \\&
    \qquad\qquad - w_1 \, (\alf u + (1-\alf) v) \, \wpot(\alf u + (1-\alf) v) \big),
\end{align*}
where we used that 
$\dpot(t) = t \wpot(t)$ 
in the last equality.
Next, because $\wpot(t)$ 
is monotonic nonincreasing for $t>0$:
%
\[
\wpot(\alf v + (1-\alf) u)
\geq
\wpot(\alf u + (1-\alf) v).
\]
Using this in the previous equation
implies that:
\begin{align*}
\dot{g}(\alf)
&\geq
C  \, \big(
w_2 \, (\alf v + (1-\alf) u) -
w_1 \, (\alf u + (1-\alf) v)
\\& =
C  \, \big(
u \, (w_2 - \alf \, (w_2 + w_1)) - 
v \, (w_1 - \alf \, (w_2 + w_1)) \big),
\end{align*}
where 
$ C \defequ (v - u) \wpot(\alf v + (1-\alf) u) > 0$.
Because $\alf < 1/2$,
the term multiplying $v$ is positive.
Using
$v < \eps \, u$
in the previous inequality yields: 
\begin{align*}
\dot{g}(\alf)
&>
C  \, \big(
u \, (w_2 - \alf \, (w_2 + w_1)) - 
\eps \, u \, (w_1 - \alf \, (w_2 + w_1)) \big)
\\ & =
C \, u \, \big(
w_2 \, (1+\alf(\eps - 1)) - 
w_1 \, (\alf + \eps(1-\alf)) \big). 
\end{align*}
If 
$w_1 = 0$ 
then we have that 
$\dot{g}(\alf) > 0$
and we can conclude that 
$g$ is increasing for 
$\alf \in (0, \frac{1}{2})$. 
If 
$w_1 > 0$ 
then we can choose 
$\eps = \frac{w_2}{w_1}$ 
to simplify the lower bound and obtain that
\begin{align*}
\dot{g}(\alf) &> 
C \, u \, \alf \, (w_2^2 - w_1^2) / w_1
> 0. \quad (\text{given that } w_2 > w_1)\nonumber
\end{align*}
Therefore, 
$g$ is increasing for 
$\alf \in (0, \frac{1}{2})$.
Proceeding analogously for the case 
$\alf \in (\frac{1}{2}, 1)$
we obtain the upper bound
\begin{align*}
\dot{g}(\alf) & < 
D \, u \, (1 - \alf)
(w_1^2 - w_2^2) / w_1
< 0, \quad (\text{given that } w_2 > w_1) 
\end{align*}
where
$D \defequ  (v - u) \wpot(\alf u + (1 - \alf)v) > 0$.
Thus, 
$g$ is decreasing for 
$\alf \in (\frac{1}{2},1)$.
We omit some of the algebraic derivations for this case,
as they are analogous to the ones in the first case.
Therefore, 
$g$ is nonconvex as it is 
increasing before $\alf = \frac{1}{2}$
and 
decreasing after $\alf = \frac{1}{2}$.
This implies that \fpotw is nonconvex which is a contradiction. This concludes the proof
for the case $r = 2$.

We now prove the general case $r > 2$ also by contradiction.
Suppose there is 
$K \in \{1,\ldots, r-1\}$
such that 
$w_K < w_{K+1}$,
and pick 
$\x \in \reals^r$
such that 
$0 < x_1 < x_2 < \ldots < x_r$.
Let 
$\y \in \reals^r$
have entries that are the same entries as \x 
except that the $(r-K)$th and $(r-K+1)$th entries are swapped,
\ie, 
$y_{r-K + 1} = x_{r-K}$ 
and
$y_{r-K} = x_{r-K + 1}$.
We choose $x_{r-K+1} < \eps \, x_{r-K}$
for some $\eps > 1$.
Under this setting we have that
\[
g(\alf) = w_K \pot(M_{\alf}) + w_{K+1} \pot(m_{\alf})
+ 
\sum_{\substack{k=1 \\ k\notin \{K, K+1\}}}^r w_k \pot(x_{r - k + 1}),
\] 
where
\begin{align*}
M_{\alf} & \defequ
\max(\alf x_{r-K} + (1-\alf)x_{r-K+1}~,~\alf x_{r-K+1} + (1-\alf)x_{r-K}),
\\
m_{\alf} & \defequ
\min(\alf x_{r-K} + (1-\alf)x_{r-K+1}~,~\alf x_{r-K+1} + (1-\alf)x_{r-K}),
\end{align*}
which implies that
\[
\dot{g}(\alf) =
w_K M_{\alf}\dpot(M_{\alf}) + w_{K+1} m_{\alf}\dpot(m_{\alf}),
\]
which can be analyzed following the same procedure
shown for the case $r = 2$.
Therefore, 
$g$ is nonconvex 
which implies that \fpotw is nonconvex,
which is a contradiction.
This concludes our proof.

\section{Proof of Proposition \ref{prop,fpotc_dif}} 
\label{ap,proof_fpot_dif}

Let $\displaystyle 0 < \delta < \min_{k \in\{1,\ldots,r\}} |x_{k} - x_{k-1}|$,
where we have defined $x_0 = 0$.
Then any vector within the ball with center at \x and radius $\delta$
has entries that are nonnegative, nonrepeated, and arranged in decreasing order.
More formally, 
\be
\y \in B_{\delta}(\x) \iff \y = \check{\y},
\ee{e,neigh_fpotw}
where
$B_{\delta}(\x) \defequ \{\y\in \reals^r ~ | ~ \norm{\x-\y} \leq \delta\}$,
with $\norm{\cdot}$ denoting the Euclidean norm.
It follows that:
\be
\fpotw(\y) = \sum_{k=1}^rw_k\pot([\check{\y}]_k)
= \sum_{k=1}^rw_k\pot([\y]_k) = \inprod{\w}{\potdot(\y)},
~ \forall \y \in B_{\delta}(\x).
\ee{e,neigh_fpotw_II}
The result follows from showing that
$\inprod{\w}{\potdot(\y)}$
is locally Lipschitz when $\y \in B_{\delta}(\x)$,
and differentiable at \x with gradient given by
$\w \odot \dpotdot{\x}$.
Showing the differentiability and gradient formula
for $\inprod{\w}{\potdot(\y)}$ is straightfoward
and follows from the differentiability of \pot.
We now show that the gradient of \fpotw is Lipschitz continuous,
which implies that \fpotw is locally Lipschitz at \x.
Let $\vu, \vv \in B_{\delta}(\x)$, then
\begin{align}
\norm{\w \odot \dpotdot{\vu} - \w \odot \dpotdot{\vv}}^2
& = \sum_{k=1}^r w_k^2 \, | \dpot(u_k) - \dpot(v_k) |^2
\nonumber\\
& \leq \norminf{\w}^2
\sum_{k=1}^r | \dpot(u_k) - \dpot(v_k) |^2
\nonumber \\
& \leq
\norminf{\w}^2 \wpot(0)^2 \norm{\vu - \vv}^2
\end{align}
as \dpot is $\wpot(0)$-Lipschitz continuous.
Therefore, the gradient of $\inprod{\w}{\potdot(\y)}$ is $L$-Lipschitz continuous
with 
$L = \norminf{\w} \wpot(0)$.

\section{Majorizers for weighted Huber-based low-rank regularizers}
\label{s,ap_weight}

A majorizer $\qpotw(\cdot~; \S, \gpotw)$
as defined in \eref{e,maj_reg_gen_weighted} should satisfy:
\begin{eqnarray}
\rpotw(\S) &=& \qpotw(\S; \S, \gpotw), \label{e,maj_a_reg_weighted}\\
\rpotw(\X) &\leq& \qpotw(\X; \S, \gpotw), \quad \forall \X\in \CMN.
\label{e,maj_b_reg_weighted}
\end{eqnarray} 
Particularly, we show that two operators,
denoted by \gpotwR and \gpotwL,
when used in \eref{e,maj_reg_gen_weighted},
lead to valid majorizers for \rpotw, \ie,
in both cases the resulting \qpotw
satisfies 
\eref{e,maj_a_reg_weighted} and \eref{e,maj_b_reg_weighted},
when \w has nondecreasing entries.
When $M<N$ (wide case),
we define \gpotwR and \gpotwL as
the following generalizations of
\eqref{e,gpot_R_wide}--%
\eqref{e,gpot_L_wide}:
\begin{eqnarray}
\gpotwR(\vs) &\defequ& (\w \odot \wpotdot{\vs}) \vone_N',
\label{e,gpot_R_wide_weighted} 
\\
\gpotwL(\vs) &\defequ& \wpot(0) \w \, \vone_N',
\label{e,gpot_L_wide_weighted} 
\end{eqnarray}
and when $M>N$ (tall case):
\begin{eqnarray}
\gpotwR(\vs) &\defequ& \vone_M (\w \odot \wpotdot{\vs})'
\label{e,gpot_R_tall_weightedl},
\\
\gpotwL(\vs) &\defequ&  \wpot(0) \vone_M \, \w'.
\label{{e,gpot_L_tall_weighted}}
\end{eqnarray}
The following theorem states that
$\qpotw(\cdot~; \S, \gpotwR)$ and $\qpotw(\cdot~; \S, \gpotwL)$
are majorizers for \rpotw at $\S \in \CMN$,
and it also provides equivalent expressions for both functions.

\begin{theorem}[Majorizers for weighted Huber-based low-rank regularizers]
\label{th,maj_Huber_weighted}

Let $\S\in \CMN$ and $\U, \V$
denote suitably sized unitary matrices
such that $\S = \U\diag{\vsig(\S)}\V'$.
Let \rpotw be defined as in \eref{e,reg_weight} with nondecreasing weights,
\ie, $0\leq w_1\leq w_2\leq \ldots\leq w_r$.
Then $\qpotw(\cdot~; \S, \gpotwR)$
and $\qpotw(\cdot~; \S, \gpotwL)$
are both majorizers for \rpotw at $\S \in \CMN$,
\ie, both satisfy \eref{e,maj_a_reg_weighted} and \eref{e,maj_b_reg_weighted}.
If $M<N$ (wide case), then 
\begin{align}
\qpotw(\X; \S, \gpotwR) &= \rpotw(\S)
+ \rinprod{\nabla \rpotw(\S)}{\X -\S}
+ \frac{1}{2}\mnormfrob{\W_{\pot, \S, \w}' \, (\X -\S)}^2,
\label{e,maj_reg_gen_R_frob_w_weighted}
\\
\qpotw(\X; \S, \gpotwL) &= \rpotw(\S) + \rinprod{\nabla \rpotw(\S)}{\X -\S}
+ \frac{1}{2} \wpot(0) \mnormfrob{\L_{\pot, \S, \w}' \, (\X -\S)}^2,
\label{e,maj_reg_gen_L_frob_w_weighted}
\end{align}
where the matrices
$\W_{\pot, \S, \w}$ and $\L_{\pot, \S, \w}$
are defined as:
\begin{align}
\W_{\pot, \S, \w} & \defequ
\U \diag{\sqrt{\w \odot (\wpotdot{\vsig(\vs)})}} \V \in\CMN.
\label{e,w_mat_weighted}
\\
\L_{\pot, \S, \w} & \defequ \U \diag{\sqrt{\w}} \V \in\CMN.
\label{e,w_mat_lips_weighted} 
\end{align}
If $M>N$ (tall case), then 
\begin{align}
\qpotw(\X; \S, \gpotwR) &= \rpotw(\S) + \rinprod{\nabla \rpotw(\S)}{\X -\S}
+ \frac{1}{2}\mnormfrob{(\X -\S) \, \T_{\pot, \S, \w}'}^2,
\label{e,maj_reg_gen_R_frob_t_weighted}
\\
\qpotw(\X; \S, \gpotwL) &= \rpotw(\S) + \rinprod{\nabla \rpotw(\S)}{\X -\S}
+ \frac{1}{2}\wpot(0)\mnormfrob{(\X -\S) \, \H_{\pot, \S, \w}'}^2,
\label{e,maj_reg_gen_L_frob_t_weighted}
\end{align}
where the matrices
$\T_{\pot, \S, \w}$ and $\H_{\pot, \S, \w}$
are defined as:
\begin{align}
\T_{\pot, \S, \w} &\defequ
\U' \diag{\sqrt{\w \odot (\wpotdot{\vsig(\vs)})}} \V' \in \CMN.
\label{e,t_mat_weighted}
\\
\H_{\pot, \S, \w} &\defequ \U' \diag{\sqrt{\w}} \V' \in \CMN.
\label{e,t_mat_lips_weighted}
\end{align}

\end{theorem}
The proof for this theorem is not provided
as it is analogous to the proof of \tref{th,maj_Huber}
in Appendix \ref{th,proof_maj_Huber}.

\section{Proof of proposition \ref{th,smooth_R_weighted}}
\label{s,ap_proof_smooth_R_weighted} 

Let $\X, \S \in \CMN$ and $\U, \V$ denote
suitable sized unitary matrices such that $\S = \U\diag{\vsig(\S)}\V'$.
Without loss of generality we assume that $M<N$ (wide case).
By \tref{th,maj_Huber_weighted} we have that:
\begin{align}
\rpotw(\X) &\leq Q_{\w}(\X; \S, \gpotwL) \nonumber 
\\
&= \rpotw(\S) + \rinprod{\nabla \rpotw(\S)}{\X -\S} \nonumber 
\\ &\quad + \frac{1}{2} \vone_M' (\gpotwL(\vsig(\S)) \odot
|\U' (\X -\S) \V|.^{\wedge}2) \vone_N.
\label{e,lips_proof_a_weighted}
\end{align}
We write the last term on the right-hand-side
of the previous inequality as follows:
\be
\frac{1}{2} \vone_M' (\gpotwL(\vsig(\S)) \odot |\U' (\X -\S) \V|.^{\wedge}2) \vone_N
=
\frac{1}{2} \wpot(0) \sum_{k=1}^M w_k \sum_{l=1}^N |[\U' (\X -\S) \V]_{kl}|^2.
\ee{e,had_lips_proof_weighted}
By replacing this in \eref{e,lips_proof_a_weighted} it follows that: 
\begin{align}
\rpotw(\X) & \leq
\rpotw(\S) + \rinprod{\nabla \rpotw(\S)}{\X -\S}
+ \frac{1}{2} \wpot(0) \sum_{k=1}^M w_k
\sum_{l=1}^N |[\U' (\X -\S) \V]_{kl}|^2
\nonumber \\ &
\leq \rpotw(\S) + \rinprod{\nabla \rpotw(\S)}{\X -\S}
+ \frac{1}{2} \norminf{\w} \wpot(0) \sum_{k=1}^M \sum_{l=1}^N
|[\U' (\X -\S) \V]_{kl}|^2
\nonumber \\ & =
\rpotw(\S) + \rinprod{\nabla \rpotw(\S)}{\X -\S}
+ \frac{1}{2} \norminf{\w} \wpot(0) \mnormfrob{\U' (\X -\S) \V}^2
\nonumber \\ & =
\rpotw(\S) + \rinprod{\nabla \rpotw(\S)}{\X -\S}
+ \frac{1}{2} \norminf{\w} \wpot(0) \mnormfrob{\X -\S}^2,
\label{e,maj_lips_proof_weighted}
\end{align}
using unitary invariance of the Frobenius norm.
Then \eref{e,maj_lips_proof_weighted} is equivalent to
$\nabla\rpotw$ being $L$-Lipschitz
where $L = \norminf{\w} \wpot(0)$
\cite[Thm.~5.8]{beck:17:fom}
when considering the inner product 
in \eref{e,inner_prod}.

\end{appendix}

\bibliographystyle{siamplain}
\bibliography{./bib-jf, ./bib-rl,./bib-jsc} 

\end{document}